\documentclass[pre,showpacs]{revtex4}

\usepackage{amsmath}
\usepackage{amsfonts}
\usepackage{amssymb}
\usepackage [ansinew] {inputenc}
\usepackage{natbib}

\usepackage{theorem}
\theorembodyfont{\upshape}
\newtheorem{postulate}{Postulate}
\newtheorem{theor}{Theorem}

\newcommand{\N} {
\ensuremath{\mathcal{N}}
}

\newcommand{\E} {
\ensuremath{\mathcal{E}}
}

\newcommand{\J} {
\ensuremath{\mathcal{J}}
}

\newcommand{\ela} {
\ensuremath{\mathcal{L}}
}

\newcommand{\ekx} {
\ensuremath{\mathcal{X}}
}

\newcommand{\open} {
\ensuremath{\mathcal{O}}
}

\begin{document}
\author{V.~Garc\'{\i}a-Morales}
\email{vladimir.garcia@uv.es}
\author{J.~Pellicer}
\title{Statistical mechanics and thermodynamics of complex systems}
\affiliation{Departament de Termodin\`{a}mica, Universitat de
Val\`{e}ncia, E-46100 Burjassot, Spain}

\begin{abstract}
\noindent
An unified thermodynamical framework based in the use of a generalized Massieu-Planck thermodynamic 
potential is proposed and a new formulation of Boltzmann-Gibbs Statistical Mechanics is established.
Under this philosophy a generalization of (classical) Boltzmann-Gibbs thermostatistics is suggested and 
connected to recent nonextensive statistics formulations. 
This is accomplished by defining a convenient squeezing function 
which restricts among the collections of Boltzmann-Gibbs configurations of the complete equilibrium closure. 
The formalism embodies Beck-Cohen superstatistics and a direct connection 
with the nonlinear kinetic theory due to Kaniadakis is provided, being the treatment presented
fully consistent with it. As an example Tsallis nonextensive statistics 
is completely rebuilt into our formulation adding new insights (zeroth law of thermodynamics,
non \emph{ad hoc} definition of the mean value of a physical quantity,...).  We relate all the
formal development to physical and measurable quantities and suggest a way to establish the
relevant statistics of any system based on determinations of temperature.
\end{abstract}
\pacs{05.20.Gg, 05.70.Ln, 05.70.Ce, 05.20.Dd}
\maketitle
\pagebreak

\section{Introduction}
\label{intro}
The physical nature of the probabilistic approach in Equilibrium Statistical Mechanics
has been discussed extensively in the literature (see \cite{Ehren, Tolman, Guttmann,
Landsberg, Gall, Lavenda}
and references therein). The entropy has long been considered as the relevant function
connecting probability theory with the thermodynamical description of the physical world.
Jaynes \cite{Jaynes} proposed in 1957 a formulation of Statistical Mechanics based on the
so-called Maximum Statistical Entropy Principle within the framework of Information Theory, by
identifying the thermodynamic entropy $S$ with the maximum value of the statistical entropy functional
$\ela(\{p_{i}\})$ ($p_{i}$ denoting the probability of an event $i$)
through the Boltzmann constant $k$. This is accomplished by means of the relation $S/k=\ela$.
Interesting criticisms of Jaynes approach have been clearly pointed out in the references (see
\cite{Guttmann}, \cite{Goldst} and references therein).

In the last two decades, there have appeared several
probabilistic views with their grounds on Jaynes approach, but applied to new definitions of
entropy. These methods have their origin in the pioneering work by Tsallis \cite{Tsallis} who
unfolded the theory of nonextensive statistics of great subsequent interest. 
This theory has been succesfully applied to a wide variety of systems and some points of the formalism 
are currently still under debate. 
Besides Ruelle's previous works (for a recent and clear discussion see, for example, \cite{Gall} and
references therein) and Hill's nanothermodynamics \cite{Hill1, Hill2}, nonextensive thermostatistics contain
new interesting views on systems that go beyond traditional thermostatistics incorporating fractality, complexity
and strong correlations. The existence of these systems are nor surprising nor exotic, and has originated a debate
on the validity of the ergodic hypothesis which has been, on the other hand, reinforced and extended after rigorous
works by other authors \cite{Gall}. There is no reason, however, to believe \emph{a priori} that all systems
in nature are ergodic. Boltzmann, in fact, encountered examples of clearly non-ergodic systems
but considered them as exceptional, unimportant, cases (the well known example of the linear chain of coupled
oscillators being an example). The study of nonequilibrium situations
and complex systems has thus motivated the apparition of new formalisms that constitute important 
efforts to face the inherent complexity of long range interactions, long term memory, fractality, etc.
There exists however some open questions and difficulties concerning nonextensive thermostatistics
as for example the definition of thermal equilibrium.

In this paper we propose a new approach to both equilibrium and nonequilibrium (stationary states)
statistical mechanics. We also address the problem of providing a sound thermodynamic basis to the statistical
methods presented which has its roots in a generalized version of the zeroth law. 
Applications of the theory to concrete physical systems and formal developments
will be given in forthcoming papers but we advance that the methods proposed 
can be applied to a wide variety of systems provided that concrete developments as Tsallis statistics, 
Beck-Cohen superstatistics and Hill's nanothermodynamics are embodied in the present formulation. 
We begin discussing equilibrium statistical mechanics
and, in the last section, we take advantage of the philosophy developed at equilibrium to extend the formalism 
to nonequilibrium situations and complex systems. Our postulates are no longer explicitly referred to the entropy 
but to a more general thermodynamic function
which attains a \emph{minimum} at equilibrium: the dimensionless characteristic thermodynamic function $\ekx$,
which coincides with $-S/k$ for an isolated system and with the Helmholtz free energy in $kT$ units
$F/kT$ for a system closed with diathermal walls in contact with a heat bath at temperature $T$.
$\ekx$ is always dependent on environment variables through the generalized partition function
which is regarded as \emph{the collection of possible (weighted) configurations
compatible with the constraints}. After introducing two postulates,
a theorem concerning the form of $\ekx$ is proven. We have, for any system
\begin{equation}
\ekx = -\ln g \label{main}
\end{equation}
where $g$ is the generalized partition function of the system. We shall no longer
use the concept of partition function, however, and we consider $g$ as the collection of configurations
within a class specified by a choice of independent variables (see below). If the system
is isolated then the collection of configurations $g$ 
coincides with the microcanonical (countable) configurations.
In this case Eq.(\ref{main}) takes the form of the celebrated expression for the
microcanonical entropy due to Boltzmann, $S=k\ln \Omega$, where $\Omega$ are the microcanonical
configurations (i.e. the total number of attainable microstates) of the system.

The condition of minimum for the appropriate thermodynamic potential for a system at equilibrium
is discussed extensively in the literature (see,
for example, Landau and Lifshitz' book \cite{LandLif}, \S 20 and \S 21). It implies that
\emph{the minimum work required to drive a system from an equilibrium
state to a neighboring one is always positive}. We make use of this point. This work
is defined as the increment of the thermodynamic characteristic function ($\equiv kT\Delta \ekx $)
when intensive variables are kept constant.
The application of the condition for a minimum in $\ekx$ leads us to the central expressions of ensemble
theory. If all attainable configurations within the classes
are allowed, Boltzmann-Gibbs (BG) equilibrium Statistical Mechanics
is derived. We consider that $\ekx$ depends on the variables chosen to characterize the system
(which we shall call environment variables) through the number of attainable configurations by the system.

In a recent paper by Vives and Planes \cite{Vives},
the importance of Massieu-Planck thermodynamic potentials in connecting Thermodynamics
with Statistical Mechanics is emphasized.
Our work shares some connections with some points of the work by these authors, 
but the approach is quite different because
the authors essentially follow the maximum entropy
principle as formulated from information theory. An advantage of our formulation,
besides the compactness achieved in the expressions (formally identical for all physical systems
whose microcanonical classes (see below) are known) is that
it provides an interface between thermodynamics and statistical mechanics
that can be easily extended to non-conventional situations as self-organized systems
or systems capable of taking decisions. In the last section we propose a scheme 
for the generalization of BG equilibrium statistics.
As a straightforward example, Tsallis non-extensive statistics \cite{Tsallis}
is reconsidered and rebuilt within this scheme
by introducing only one additional assumption. 

The formulation presented here is general and applies also to small systems (with the only exception
of Section \ref{fluctu} which deals with fluctuations in macroscopic systems) when homogeneity
of $\ekx$ is not demanded. We summarize below some features and issues of the formulation 
that could be of general interest:

1. Compactness and generality in the expressions derived. Only the knowledge of the
dimensionless characteristic thermodynamic function $\ekx$
in terms of generalized extensive variables $X_{j}$ and intensive ones $y_{i}$
are required to provide a complete
thermodynamic description of \emph{any} system. If the configurations $g$ are also
considered in terms of these variables an statistical framework can be developed.

2. Hill's nanothermodynamics \cite{Hill1, Hill2} is embodied in the approach followed. This formalism 
(developed previously) is, besides Gross microcanonical thermostatistics \cite{Gross}
(not discussed here), a quite convenient framework to explain the nonextensive (equilibrium) thermodynamics
of systems with size comparable to the range of the forces involved. It is important to make here
a distinction between an extensive or nonextensive thermodynamics and an extensive or nonextensive
statistics. \emph{A nonextensive statistics does not necessarily imply a nonextensive thermodynamics.}
Importantly, Hill's formalism makes stable the previous classical development of 
Thermodynamics to the domain where fluctuations in physical quantities are important.

4. The statistical formulation can be easily extended and applied 
to non conventional situations (complex and self-organized systems,
nonequilibrium and metaequilibrium, etc.) The method is general and it is not restricted, in principle,
to some concrete choices made below to illustrate how it works.

5. In particular, Tsallis statistical mechanics is rebuilt
within our framework. New insights concerning this theory are provided. Concretely, 
a new definition for the mean value of a mechanical quantity is suggested. 
This definition is different to those encountered in the foundational
papers of nonextensive statistics (see below) \emph{and is not given ad hoc} but arises naturally from the formalism.
Tsallis form for the entropy in terms of probabilities is derived.

5. The generalized formulation presented can also be related easily to recent developments
on Statistical Mechanics. We discuss, concretely, Beck-Cohen superstatistics.

6. The formulation is fully consistent with the kinetic theory previously proposed by Kaniadakis
which generalizes the Boltzmann kinetic theory and the Fokker-Planck picture. 
Besides this previous work, irreversibility can be understood
in a broader context. The close interrelation between both formalisms makes the formulation presented well
grounded kinetically and can thus serve as a basis for the study of a wide variety of statistical systems.

The paper is organized as follows: in Section \ref{charf} the dimensionless
characteristic thermodynamic function $\ekx$
is introduced from thermodynamical considerations showing how the knowledge of this function
provides the complete thermodynamical description of the system.
In Section \ref{GB} two postulates on the dimensionless characteristic function are presented,
relating it to the configurations that a physical system can attain. An important theorem on the
form of the function $\ekx$ is also proven.
In Section \ref{ET} ensemble theory is developed over the two preceding postulates.
In Section \ref{prob} probability is introduced and the Gibbs-Shannon form for the entropy is derived.
In Section \ref{fluctu} a general theory of fluctuations is given. In Section \ref{Ex},
the previous formalism is applied to the usual ensembles.
Finally, in Section \ref{generalization}, it is shown in general how
BG equilibrium statistics can be extended
to complex systems and nonequilibrium. Squeezed statistical mechanics is formulated and a connection
with the nonlinear kinetic theory by Kaniadakis, which provides a generalization of the celebrated 
Boltzmann kinetic equation, is established. In Section \ref{app} 
Tsallis nonextensive statistics is discussed and completely rebuilt as a
concrete formal application by introducing an appropiate squeezing function \emph{motivated from physical grounds}. 
New insights in Tsallis statistics are presented. The connection with Beck-Cohen superstatistics is also established. 

We remark that, although Sections \ref{charf} to \ref{fluctu} are largely conventional (as are referred to 
classical BG equilibrium thermostatistics) the constructive procedure 
that we use to develop the non conventional statistics (Section \ref{generalization}) is presented. 

\section{Formulation of BG thermostatistics}
\subsection{The characteristic thermodynamic function as a thermodynamic potential}
\label{charf}

For any given system, the following differential form for the entropy $\J=S/k$ holds
\begin{equation}
d\J-\sum_{l}y_{l}dX_{l} = 0 \label{generdE}
\end{equation}
Here $X_{l}$ is an extensive variable and $y_{l}$ is its conjugate intensive one divided
by $kT$. For a simple pure substance under hydrostatic conditions $\{X_{l}\}=\{E,V,N\}$
and $\{y_{l}\}=\{1/kT,p/kT,-\mu /kT\}$ are the two sets of variables and Eq.(\ref{generdE})
(where $E$, $V$, $N$, $p$ and $\mu$  are the energy, volume, number of particles, pressure and
chemical potential, respectively) can be written as
\begin{equation}
\frac{1}{k}dS-\frac{1}{kT}dE-\frac{p}{kT}dV+\frac{\mu}{kT}dN = 0\label{concdE}
\end{equation}
In Eq.(\ref{concdE}) the differential of the entropy is expressed in terms of the differentials
of $E$, $V$ and $N$ being this the adequate form (for practical purposes) to describe the thermodynamics
of a completely closed system (as it is characterized by the specification of
these three extensive variables). However,
for a partially open system, other choices of variables are more adequate. The variables
chosen to characterize the system are called \emph{environment} (or \emph{natural}) variables.

Let us consider the sets of variables $\{X_{l}\}$ and $\{y_{l}\}$. Each of them
can be separated into two subsets satisfying $\{X_{l}\}=\{X_{j}\} \cup \{X_{i}\}$ and
$\{y_{l}\}=\{y_{j}\} \cup \{y_{i}\}$. We define $\{X_{j}\}$ as the subset of extensive
environment variables and $\{y_{i}\}$ as the subset of intensive environment variables.
The subsets $\{X_{i}\}$ and $\{y_{j}\}$ contain the conjugate variables
(which are not environment variables)
of those within the former subsets. Thus, an environment extensive
variable $X_{j}$ is the conjugate of an intensive non-environment one $y_{j}$, while an
intensive environment variable $y_{i}$ is the conjugate of an extensive non-environment one
$X_{i}$. Hence, the complete set of environment variables is $\{X_{j}\} \cup \{y_{i}\}$.
For an isolated system $\{y_{i}\}=\{\emptyset \}$ and, therefore, $\{X_{i}\}=\{\emptyset \}$.

For a specification of $M$ environment variables,
if the subset $\{X_{j}\}$ contains $m$ variables, $\{y_{i}\}$ contains $M-m$ variables
and Eq. (\ref{generdE}) can be rewritten as
\begin{equation}
d\J-\sum_{j=1}^{m}y_{j}dX_{j}-\sum_{i=m+1}^{M}y_{i}dX_{i} = 0\label{generdE2}
\end{equation}
If we add and substract a term $d\left(\sum_{i=m+1}^{M}y_{i}X_{i}\right)$, we obtain
\begin{equation}
d\left(\J-\sum_{i=m+1}^{M}y_{i}X_{i}\right)-\sum_{j=1}^{m}y_{j}dX_{j}+
\sum_{i=m+1}^{M}X_{i}dy_{i} = 0\label{generdE3}
\end{equation}
We define the dimensionless characteristic thermodynamic function $\ekx$ as
\begin{equation}
\ekx  \equiv \sum_{i=m+1}^{M}y_{i}X_{i}-\J \label{ekk}
\end{equation}
We adopt henceforth Einstein summation convention for repeated indexes
(and then $\ekx  \equiv y_{i}X_{i}-\J$).
From Eq.(\ref{generdE3}) we thus have
\begin{equation}
d\ekx=-y_{j}dX_{j}+ X_{i}dy_{i} \label{genED}
\end{equation}
This differential equation can be considered as
a generalized form of Eq.(\ref{generdE}) which was written
for a completely closed system. As we have seen above, for an isolated system
$\{X_{j}\}=\{X_{l}\}$, $\{y_{j}\}=\{y_{l}\}$, $\{X_{i}\}=\{\emptyset\}$, $\{y_{i}\}=\{\emptyset \}$,
and, from Eq. (\ref{ekk}), $\ekx=-\J$. With these considerations, Eq.(\ref{generdE})
is recovered. It is important emphasizing here that Eq.(\ref{genED}) is valid for any system
independently of its constraints. This equation also shows an important feature of the
dimensionless characteristic function: given the set of environment variables, the knowledge
of $\ekx$ in terms of these allows the complete thermodynamical description of the system.
Non-environment variables $y_{k}$
($y_{k} \in \{y_{j}\}$) and $X_{s}$ ($X_{s} \in \{X_{i}\}$)
are then specified through the following relations
\begin{eqnarray}
y_{k}&=& -\left(\frac{\partial \ekx}{\partial X_{k}} \right)_{\{X_{j}\} j \neq k,\{y_{i}\}} \label{pfuncekx} \\
X_{s}&=&\left(\frac{\partial \ekx}{\partial y_{s}} \right)_{\{X_{j}\},\{y_{i}\} i \neq s} \label{Qfuncekx}
\end{eqnarray}
The dimensionless characteristic function $\ekx$ is a generalized Massieu-Planck function.
Massieu-Planck functions
are entropic thermodynamic potentials defined as Legendre transformations of the entropy
\cite{Vives}. For the concrete cases of an isothermal closed system and an isothermal-isobaric
impermeable system, the dimensionless characteristic function defined by Eq.(\ref{ekk}) coincides
with the thermodynamic potentials first introduced by Massieu \cite{Massieu} and Planck \cite{Planck},
respectively through a minus sign.

In macroscopic systems $\ekx$ is an homogeneous function of first order in extensive environment
variables and of zero order in intensive environment variables.
Hence, due to the linearity of the differential form of $\ekx$, Eq. (\ref{genED}), the following relation
holds
\begin{equation}
\ekx  = -y_{j}X_{j} \label{Euler}
\end{equation}
for any macroscopic system. Eq.(\ref{Euler}) is the well known Euler's equation for a macroscopic system.
As a consequence of it, by taking the differential in each side of Eq.(\ref{Euler}), and comparing
the result with Eq.(\ref{genED}) we obtain the Gibbs-Duhem equation for macroscopic systems
\begin{equation}
X_{j}dy_{j}+ X_{i}dy_{i}=0
\label{GibbsDuhem}
\end{equation}

One can relax, in general, the requirement of $\ekx$ to be an homogeneous function of first order
in extensive environment variables. This is the case of small systems where this homogeneity
breaks. An outstanding feature of these systems is the presence of finite-size effects due to
forces of range comparable to the size of the system and to its coupling to the environment.
Hill's nanothermodynamics \cite{Hill1, Hill2} is an important approach
to explain their thermal properties. Here we follow a different approach
which proves to be equivalent to Hill considerations.
We suggest that for small systems the deviations from the behavior
of macroscopic systems in the thermodynamic limit can be accounted for by introducing
a term $\open$ in Eq.(\ref{Euler}) by the following way
\begin{equation}
\ekx  = -y_{j}X_{j}-\open \label{Hillekx}
\end{equation}
The Gibbs-Duhem equation takes now the form
\begin{equation}
d\open=-X_{j}dy_{j}-X_{i}dy_{i}
\label{HillGibbsDuhem}
\end{equation}
These considerations are consistent with Hill's definition of the \emph{subdivision potential} $\E$
\cite{Hill1, Hill2}. In fact $\open \equiv -\E/kT$. Note that within this way of introducing
the inhomogeneities, formal symmetries appear in the equations of thermodynamics among
variables chosen to characterize the system. On one hand Eqs.(\ref{generdE}) and
(\ref{HillGibbsDuhem})
\begin{eqnarray}
d\J=y_{l}dX_{l} \label{SgenerdE} \\
d\open=-X_{l}dy_{l} \label{SHgenerdE}
\end{eqnarray}
and on the other Eqs.(\ref{ekk}) and (\ref{Hillekx})
\begin{equation}
\ekx=y_{i}X_{i}-\J=-y_{j}X_{j}-\open \label{Sekx} \\
\end{equation}
show that there exists two different entropic functions $\open$ and $\J$ that can lead, through the convenient
Legendre transforms, to the same thermodynamic description of a system, specified by the
dimensionless characteristic thermodynamic function $\ekx$. The dimensionless entropy $\J$
corresponds to a thermodynamic description in which the independent variables are all extensive.
In the case of a description based on $\open$, the independent variables are all intensive.
For a partially open system, the appropriate function is $\ekx$. The conventional entropy
of the system is obtained in terms of $\ekx$ by closing completely the system through
the adequate Legendre transforms while $\open$ is obtained opening completely
the system by making the Legendre transforms in the reverse order. 

The results of this section have shown that the knowledge of $\ekx$ in terms of the
environment variables is enough to completely describe the thermodynamical behavior of a system.
In the following section $\ekx$ becomes the central function in connecting the microscopic
``counting over configurations" to the macroscopic world. It is important to remark at this point
that thermodynamical non environment variables constitute observed values of microscopic averages. 
Establishing the form that these microscopic averages take is one of the objectives of the following sections.
If we denote by \emph{obs} thermodynamically observed variables, Eq. (\ref{ekx}) can be rewritten as 
\begin{equation}
\ekx  \equiv y_{i}X_{i,obs}-\J \label{ekx}
\end{equation}

\subsection{Postulates of Equilibrium Statistical Mechanics}
\label{GB}

A physical system is found, in a given instant of time, to be in a particular configuration.
Constraints imposed to the system limit the number of configurations that it can adopt.
When some of these constraints are removed the system evolves through an irreversible process
occupying new possible configurations until equilibrium
(where new but less restrictive constraints are specified) is reached. The result is that
the total number of configurations has been increased attaining a maximum
on going to equilibrium, where the function $\ekx$ attains a minimum.
These facts always hold for an irreversible process. A larger
number of attainable configurations is related to a lower value of $\ekx$. Furthermore,
each configuration contains a partial thermodynamic description in itself: it is compatible
with the specification of some variables. We can then introduce the concept of \emph{collections or classes
of configurations} $g_{\{X...\}\cup \{y...\}}$
to denote the sets of configurations compatible with the
specification of a set of variables.
On collecting these classes of configurations we obtain a complete thermodynamic description of the system
if we relate them to a thermodynamic potential. The dimensionless characteristic
thermodynamic function $\ekx$ is the most general form for an entropic thermodynamic potential,
embodying in itself all the choices of concrete entropic thermodynamic potentials that can
be done, compatible with specifying a set of environment variables.

For an irreversible process
at constant intensive environment variables we have, from Eq.(\ref{ekx})
\begin{equation}
\Delta \ekx=y_{i}\Delta X_{i}-\Delta\J \le 0 \label{incr}
\end{equation}
for the system on going to equilibrium. Eq.(\ref{incr}) implies that
it is necessary to do a positive work on a system to drive it from equilibrium to
a nonequilibrium neighboring state \cite{LandLif}. This consideration applies,
in principle, to systems with many particles, as it is of thermodynamic character.
It is not obvious to hold for systems with few particles due to the microscopic
reversibility of the dynamical behaviour \cite{Lebow}. However, it constitutes a good starting point
for all what follows since it assures the very desirable property of 
stability (convexity in the free energy, concavity in the entropy) that can be extended
later to other situations. This \emph{demanded} convexity in the free energy is not a matter
for worrying about when studying, for example, phase transitions of small systems with the
apparition of convex intruders in the entropy, since we supplement the information concerning small systems
(see below) through the Hill's subdivision thermodynamic potential (the theory
for this important concept can be found elsewhere \cite{Hill1, Hill2}).

It is useful to consider in more detail the concept of classes of configurations. At equilibrium
these classes can be hierarchically interrelated and defined, the classes being unions of subclasses. 
In general, the most important class 
is what we call below \emph{the characteristic class} of the system: the collection of elements compatible
with the specification of environment variables $\{X_{j}\}$ and $\{y_{i}\}$, $g_{\{X_{j}\} \cup \{y_{i}\}}$. 
Another important class is that of the total number of microstates of the objective system, 
the \emph{microcanonical class} $g_{\{X_{j}\}}$ in which only the extensive variables of the system are specified. For an isolated system, the
microcanonical class is also the characteristic one. In equilibrium, all the classes and interrelations between
them can be established, and the main classes described above can be decomposed in subclasses through the following way
\begin{eqnarray}
g_{\{X_{j}\}}&=&\bigcup_{\{X_{i}\}}g_{\{X_{j}\} \cup \{X_{i}\}}
=\sum_{\{X_{i}\}}g_{\{X_{j}\} \cup \{X_{i}\}} \label{mic0} \\
g_{\{X_{j}\} \cup \{y_{i}\}}&=&\bigcup_{\{X_{i}\}}g_{\{X_{j}\} \cup \{y_{i}\} \cup \{X_{i}\}}
=\sum_{\{X_{i}\}}g_{\{X_{j}\} \cup \{y_{i}\} \cup \{X_{i}\}} \label{mic} 
\end{eqnarray}
where the sums are over all the possible specifications of each of the variables belonging to the set $\{X_{i}\}$.
The microcanonical subclasses $g_{\{X_{j}\} \cup \{X_{i}\}}$ are particularly interesting since they constitute the so-called 
\emph{degeneracy} of a configuration, and can be evaluated directly from the Hamiltonian after specifying variables
$\{X_{j}\}$ and $\{X_{i}\}$. Knowing how the characteristic class is decomposed in microcanonical subclasses is,
hence, mandatory and, stated in other way, it constitutes the celebrated Gibbs theorem. We present below a new derivation of this
theorem that has a certain (vague) resemblance with the foundations of statistical mechanics by the method of counting
in the microcanonical ensemble due to Balian and Balazs \cite{BB}. We choose, however, to focus the discussion in the 
concept of classes of configurations, in which a system can be, instead of probability. This is appropiate 
when studying the important issue of irreversibility since microcanonical classes evolve with time, a feature which is absent in probability distributions \cite{Goldst}.   

An irreversible process can be pictured as the result
of removing one or more constraints. Removing these implies enlarging the
collection of configurations as the classes of configurations are formed 
by elements compatible with the constraints and for less restrictive
constraints there exists more elements compatible with them. It is
important to note here that this statement holds for systems
which are uncapable of restricting (or enlarging) by themselves their collections
of configurations. The role of the constraints is crucial (and the only
essential thing) for these systems
in constructing the classes. In the last section of this paper
we study complex, organized systems that can select among restricted
(or enlarged) classes of configurations. Let us study first the case of systems that have not this capability, i.e.
Boltzmann-Gibbs (and, as we shall see, ergodic) systems. We postulate:

\begin{postulate}
\label{p1}
For any physical system there exists a dimensionless characteristic thermodynamic
potential $\ekx$ which is a monotonously decreasing function of the size of 
the characteristic collection of configurations of the system 
(abbreviated by $g$) i.e. $\ekx(g) \le \ekx(g+|\Delta g|)$. $g$ coincides with the numerical value 
of the appropiate partition function. Therefore, when the system is isolated $g$ coincides
with the total number of microstates (the microcanonical class).
\end{postulate}

\begin{postulate}
\label{p2}
The dimensionless characteristic thermodynamic potential is an extensive quantity,
i.e. for two independent systems $A$ and $B$, $\ekx_{A+B}=\ekx_{A}+\ekx_{B}$.
\end{postulate}

These two postulates allow us to build the whole formalism of ensemble theory and fluctuations.
This is shown in the following sections. We prove first two theorems on the form
of the dimensionless characteristic thermodynamic function.

\begin{theor}
\label{t1}
For $\N$ independent identical systems, each of them with $g$ elements $\ekx$ satisfies $\ekx (g^\N)=\N \ekx (g)$.
\end{theor}

The proof of this theorem is immediate if one considers that, if the systems are independent,
the number of elements of the whole ensemble are given by $g^{\N}$. Let
$\ekx_{\N}$ be the dimensionless characteristic function of the ensemble and $\ekx_{1}$
that of one system. Then, by Postulate \ref{p2}, we have $\ekx_{\N}=\ekx (g^{\N})=\N \ekx_{1}=\N \ekx (g)$.

\begin{theor}
\label{t2}
The dimensionless characteristic thermodynamic function is given by
$\ekx (g)= -|\lambda| \ln g$ where $\lambda$ is a dimensionless constant.
\end{theor}

In the proof of this theorem we follow an analogous treatment to the developed by Landsberg
\cite{Landsberg} for the entropy (pp. 126-128 of his book). Let $r$, $g$, $n$ be arbitrary
positive integers, with $m$ determined by the following chain of inequalities
\begin{equation}
r^{m} \le g^{n} \le r^{m+1}
\end{equation}
Forming $\ln x$ for each term of this equation, one finds
\begin{equation}
m \ln r \le n \ln g \le (m+1)\ln r
\end{equation}
whence
\begin{equation}
\frac{m}{n}  \le \frac{\ln g}{\ln r} \le \frac{m+1}{n}
\end{equation}
An exactly similar relation is satisfied by $\ekx$, by virtue of Theorem \ref{t1}
\begin{equation}
\frac{m}{n}  \le \frac{\ekx (g)}{\ekx (r)} \le \frac{m+1}{n}
\end{equation}
It follows that
\begin{equation}
\left|\frac{\ekx (g)}{\ekx (r)}- \frac{\ln g}{\ln r} \right| \le \frac{1}{n}
\label{goodp}
\end{equation}
The l.h.s. of Eq.(\ref{goodp}) is independent of $m$, while $n$ is arbitrary. Hence,
upon taking $n$ arbitrarily large, it is obtained that
\begin{equation}
\frac{\ekx (g)}{\ln g}=\frac{\ekx (r)}{\ln r}
\end{equation}
is a constant, $\lambda$, independent of $r$. Then
\begin{equation}
\ekx (g)=\lambda \ln g
\end{equation}
From Postulate \ref{p1} we see that $\lambda \le 0$ and, hence, we obtain the desired
result.

$|\lambda|$ is a dimensionless constant which
adjust the scale of temperatures. For example,
for a mean number $<N>$ of molecules of ideal gas occupying a macroscopic volume $V$
of an isothermal open region one finds $\ekx=-pV/kT=\lambda <N>$.
The equation of state of the ideal gases can be reproduced empirically and the fitting of
the experimental curves leads to an expression $pV=<N>kT$ for the ideal gases. Then,
$\lambda$ can be conveniently identified with a numerical value of $-1$ and, hence,
we can write
\begin{equation}
\ekx (g)= -\ln g  \label{hmain}
\end{equation}
We shall use this result in all what follows. 

\subsection{Ensemble Theory}
\label{ET}

The three fundamental expressions obtained for $\ekx$,
Eqs. (\ref{ekx}), (\ref{genED}) and (\ref{hmain})
contain, expressed in compact form, the main expressions of formal
ensemble theory of equilibrium statistical mechanics and thermodynamics.
A workable version
of the former in the present framework is necessary and this is the aim
of this Section. Eq.(\ref{hmain}) can be rewritten,
in terms of the specified environment variables, as
\begin{equation}
\ekx = -\ln g_{\{X_{j}\} \cup \{y_{i}\}}  \label{hmain2}
\end{equation}
The main advantage of focusing the discussion in the function $\ekx$ is that all equations
obtained are valid for any system.
Let us consider a completely isolated system. All environment
variables for this system are extensive and these specify a total number of
configurations $g_{\{X_{j}\}} = g_{\{X_{l}\}} \equiv \Omega$
compatible with the constraints imposed.
For this system the dimensionless characteristic thermodynamic function $\ekx$
coincides with $-\J$, see Eq.(\ref{ekx}). The (dimensionless) entropy
then can be defined as minus the microcanonical dimensionless characteristic function.
This definition of the entropy is general and valid also for partially opened systems
if one always considers the whole number of configurations compatible with regarding the system
as completely closed. As for a partially opened system there are different ways of closing it
\emph{all configurations compatible with the different ways of closing the system should be counted
and added}. In the following we consider a discrete spectrum of configurations for simplicity.
Analytically we can write for the entropy $\J$ in general
\begin{equation}
\J=\ln \sum_{\{X_{i}\}}g_{\{X_{j}\} \cup \{X_{i}\}} = \ln g_{\{X_{j}\}}
\label{entro2}
\end{equation}
where the property stablished in Eq. (\ref{mic0}) has been used. Note that this definition
is referred to the total number of configurations
compatible with extensive environment variables. The very meaning of closing a system
is specifying a concrete value for the extensive variables.
The set of extensive environment
variables change for different systems, but the definition of the entropy
is always formally identical.
A constraint is imposed, in general, when a thermodynamic variable
is forced to have a concrete value. As indicated above, microcanonical classes
are directly related to the Hamiltonian and, hence, it is convenient
to know the combination of classes of configurations $g_{\{X_{j}\} \cup \{X_{i}\}}$
on which $g_{\{X_{j}\} \cup \{y_{i}\}}$ is decomposed. 

We have, from Eqs.(\ref{hmain2}) and (\ref{mic})
\begin{equation}
\ekx=-\ln g_{\{X_{j}\} \cup \{y_{i}\}}= -\ln \sum_{\{X_{i}\}}
g_{\{X_{j}\} \cup \{y_{i}\} \cup \{X_{i}\}}
\label{ekx2}
\end{equation}

If we specify a concrete value for the extensive non environment variables  
we have, by using Eqs. (\ref{ekx}) and (\ref{entro2})
\begin{eqnarray}
\ekx_{\{X_{i}\}}&=& y_{i}X_{i}-\J _{\{X_{i}\}}=
y_{i}X_{i}-\ln g_{\{X_{j}\} \cup \{X_{i}\}} \nonumber \\
&=&-\ln \left[g_{\{X_{j}\} \cup \{X_{i}\}}e^{-y_{i}X_{i}}\right]
\label{ekx4}
\end{eqnarray}
where $X_{i,obs}=X_{i}$ has been used in Eq.(\ref{ekx}) since this variable 
is now specified. From Eq.(\ref{ekx2}), we can write
\begin{equation}
\ekx_{\{X_{i}\}}=-\ln g_{\{X_{j}\} \cup \{y_{i}\} \cup \{X_{i}\}}
\label{ekx5}
\end{equation}
and, by comparing Eqs.(\ref{ekx4}) and (\ref{ekx5}) we obtain
\begin{equation}
g_{\{X_{j}\} \cup \{y_{i}\} \cup \{X_{i}\}}=
g_{\{X_{j}\} \cup \{X_{i}\}}e^{-y_{i}X_{i}}
\label{ensem}
\end{equation}
and therefore, from Eq.(\ref{ekx2})
\begin{equation}
\ekx=-\ln \sum_{\{X_{i}\}}g_{\{X_{j}\} \cup \{X_{i}\}}e^{-y_{i}X_{i}}
\end{equation}
Hence, Gibbs theorem is proven and ensemble theory is thus formulated through the previous approach.
The arguments presented above have
lead us to general expressions obeyed for all systems in Boltzmann-Gibbs equilibrium.
We summarize below the most important equations obtained:
\begin{eqnarray}
\ekx &=& -\ln g_{\{X_{j}\} \cup \{y_{i}\}} \nonumber \\
&=&-\ln \sum_{\{X_{i}\}}g_{\{X_{j}\} \cup \{X_{i}\}}e^{-y_{i}X_{i}} \label{genpar} \\
d\ekx &=&-y_{j,obs}dX_{j}+ X_{i,obs}dy_{i} \label{dgenpot} \\
\J &=& -\ekx+y_{i}X_{i,obs} \nonumber \\
&=&-\ekx+y_{i}\left(\frac{\partial \ekx}{\partial y_{i}} \right)_{\{X_{j}\},\{y_{i^{*}}\}} \nonumber \\
&=& \ln g_{\{X_{j}\}} \label{genent} \\
\J _{\{X_{i}\}} &=& -\ekx_{\{X_{i}\}}+ y_{i}X_{i} \label{genentp}
\end{eqnarray}
The subindex $i^{*}$ makes reference to all variables belonging to the set $\{y_{i}\}$ excluding only that
involved in the derivative.
Note that in Eq.(\ref{genpar}) \emph{the sum is carried over the whole spectrum of values
that extensive non-environment variables can take, as these are allowed to vary freely}.
Thus, Eq.(\ref{genpar}) serves as a definition of $\ekx$. The argument in the logarithm is the generalized
partition function. Eq.(\ref{dgenpot}) allows the thermodynamical description of the system
and Eq.(\ref{genent}) is a definition of entropy as a function of $\ekx$. Finally Eq.(\ref{genentp})
is a useful definition of the entropy when concrete classes of configurations compatible with
the specification of non-environment extensive variables are taken into account.
For the entropy $\open$ we have
\begin{eqnarray}
\open &=&
-\ekx-y_{j,obs}X_{j} \nonumber \\
&=& -\ekx+X_{j}\left(\frac{\partial \ekx}{\partial X_{j}} \right)_{\{X_{j^{*}}\},\{y_{i}\}}
\label{Hillgenent} \\
\open _{\{y_{j}\}}&=&-\ekx_{\{y_{j}\}}-y_{j}X_{j} \label{Hillgenentp}
\end{eqnarray}
For the case of a completely open system one has $\ekx=-\open$ and, hence, from Eq.(\ref{genpar})
\begin{equation}
\open=\ln \sum_{\{X_{i}\}}g_{\{X_{i}\}}e^{-y_{i}X_{i}} \qquad \{y_{i}\}=\{y_{l}\}
\end{equation}
and so, in general, for any system we have
\begin{equation}
\open=\ln \sum_{\{X_{i}\}}g_{\{X_{i}\}}e^{-y_{i}X_{i}} \label{genopen}
\end{equation}
In the case of macroscopic systems, $\open = \open _{\{y_{j}\}}=0$.
All above equations can be supplemented to the following, relating the number of characteristic
elements of a system to the \emph{countable} microstates resulting from specifying extensive variables.
\begin{equation}
g_{\{X_{j}\} \cup \{y_{i}\}}= \sum_{\{X_{i}\}}
g_{\{X_{j}\} \cup \{X_{i}\}}e^{-y_{i}X_{i}}
\label{micr}
\end{equation}
We have deal with discrete specifications of the elements within a class. If a continuous spectrum
is considered, the last equation must be written as
\begin{equation}
g_{\{X_{j}\} \cup \{y_{i}\}}= \int_{\{X_{i}\}}
g_{\{X_{j}\} \cup \{X_{i}\}}e^{-y_{i}X_{i}} \prod_{i}\frac{dX_{i}}{\Delta_{Xi}}
\label{micrcont}
\end{equation}
We see that in the case of a continuous spectrum
the Laplace transform is the appropriate mathematical tool
for passing from one ensemble to another with less extensive constraints.
The reverse process can then be mathematically described in terms
of the inverse Laplace transform. By using the
Bromwich inversion formula \cite{Pathria}, we can write
\begin{equation}
g_{\{X_{j}\} \cup \{X_{i}\}}= \frac{1}{2\pi i}\int_{\gamma_{\{y_{i}\}}-i\infty}^{\gamma_{\{y_{i}\}}+i\infty }
g_{\{X_{j}\} \cup \{y_{i}\}}e^{y_{i}X_{i}} \prod_{i}\frac{dy_{i}}{\Delta_{yi}}
\label{micrcontinv}
\end{equation}
where $\gamma_{\{y_{i}\}}$
is chosen so that the poles and singularities of the integrand in the complex plane are at the left
of the integration path for each of the variables $\{y_{i}\}$. For small systems,
however, these transforms become non-local leading to ensemble non-equivalence.
Eqs.(\ref{genpar}), (\ref{dgenpot}), (\ref{genent}), (\ref{genentp}) and (\ref{genopen})
are the most important expressions of Boltzmann-Gibbs equilibrium thermostatistics.

\subsection{Probability}
\label{prob}

From the previous treatment we can define now the probability $\mathcal{P}$
of measuring values for extensive non-environment variables $\{X_{s}\}$
($\{X_{s}\} \subset \{X_{i}\}$) as the chance of the system for being
in a given configuration compatible with the specification of these variables. This is
simply the quotient of the numerical value of configurations at a specified value
of variables $\{X_{s}\}$, $g_{\{X_{j}\} \cup \{y_{i}\} \cup \{X_{s}\}}$ to the value of the whole collection
\begin{eqnarray}
\mathcal{P}_{\{X_{s}\}}&=&\frac{g_{\{X_{j}\} \cup \{y_{i}\} \cup \{X_{s}\}}}
{g _{\{X_{j}\}\cup \{y_{i}\}}}=
\frac{g_{\{X_{j}\} \cup \{y_{i}\} \cup \{X_{s}\}}}
{\sum_{\{X_{s}\}}g_{\{X_{j}\} \cup \{y_{i}\} \cup \{X_{s}\}}} \nonumber \\
&=&\frac{g_{\{X_{j}\} \cup \{X_{s}\}}e^{-y_{s}X_{s}}}
{\sum_{\{X_{s}\}}g_{\{X_{j}\} \cup \{X_{s}\}}e^{-y_{s}X_{s}}}
\label{proba}
\end{eqnarray}
From Eqs.(\ref{ekx2}) and (\ref{ekx5}) we obtain
\begin{equation}
\mathcal{P}_{\{X_{s}\}}=e^{\ekx-\ekx_{\{X_{s}\}}}
\label{px}
\end{equation}
The probability distribution is normalized, as is evident from Eq.(\ref{proba}).
\begin{equation}
\sum_{\{X_{s}\}}e^{\ekx-\ekx_{\{X_{s}\}}}=1
\label{pnormal}
\end{equation}

The values of non-environment variables fluctuate
as these are not specified \emph{a priori}. The observed macroscopic values
for non-environment variables are given by the differential equation (\ref{dgenpot})
obtained from thermodynamical considerations. From that equation, we have,
for a variable belonging to the set $\{X_{i}\}$,
\begin{equation}
X_{s, obs}=\left(\frac{\partial \ekx}{\partial y_{s}} \right)_{\{X_{j}\},\{y_{i}\} i \neq s} \label{der}
\end{equation}
From Eq.(\ref{der}) and using Eqs.(\ref{ensem}) and (\ref{genpar}) and the definition of probability
given above, Eq.(\ref{proba}), a straightforward calculus leads to
\begin{equation}
X_{s,obs}=\frac{\sum_{X_{s}}\left[X_{s} \mathcal{P}_{\{X_{s}\}} \right]}{\sum_{X_{s}}\mathcal{P}_{\{X_{s}\}}} = <X_{s}>
\label{observ}
\end{equation}
i.e. the observed value of the variable $X_{s}$ is the ensemble average of this quantity
$<X_{s}>$ over all attainable
configurations. If the observed values are related to time averages, Eq. (\ref{observ})
constitutes the so-called \emph{ergodic hypothesis} which is of fundamental importance for the
probabilistic approach in Boltzmann equilibrium statistical mechanics.

The probability $p_{k,\{X_{s}\}}$ for the system to be \emph{in a concrete configuration} $k$
compatible with the specification of the constraints and variables $\{X_{s}\}$
($\{X_{s}\} \subset \{X_{i}\}$) can now be defined as
\begin{eqnarray}
p_{k,\{X_{s}\}}&=&\frac{\mathcal{P}_{\{X_{s}\}}}{g _{\{X_{j}\}\cup \{X_{s}\}}}
=\frac{g_{\{X_{j}\} \cup \{y_{i}\} \cup \{X_{s}\}}}
{g _{\{X_{j}\}\cup \{X_{s}\}}g _{\{X_{j}\}\cup \{y_{i}\}}}\nonumber \\
&=& \frac{e^{-y_{s}X_{s}}}
{\sum_{\{X_{s}\}}g_{\{X_{j}\} \cup \{X_{s}\}}e^{-y_{s}X_{s}}} \nonumber \\
&=& \frac{e^{-y_{s}X_{s}}}
{\sum_{k} e^{-y_{s}X_{s}|_{k}}}
\label{probb}
\end{eqnarray}
In terms of this probability distribution, the dimensionless characteristic thermodynamic
function, Eq.(\ref{genpar}), can be rewritten as
\begin{equation}
\ekx=-\ln \sum_{k}e^{-y_{i}X_{i}|_{k}} \label{simgenpar}
\end{equation}
From Eqs.(\ref{entro2}) and (\ref{px}) we obtain
\begin{equation}
p_{k,\{X_{s}\}}=e^{\ekx-\ekx_{\{X_{s}\}}-\J_{\{X_{s}\}}}
\label{px2}
\end{equation}
which is also normalized leading trivially to the same result
concerning observable values,
Eq.(\ref{observ}).
For the entropy, we have,
by using Eqs.(\ref{genpar}) and (\ref{genent})
\begin{eqnarray}
\J&=&y_{i}X_{i,obs}-\ekx=\sum_{k}p_{k}\left(y_{i}X_{i}-\ekx\right)_{k}=\nonumber \\
&=&\sum_{k}p_{k}\ln
\frac{\sum_{k} e^{-y_{i}X_{i}|_{k}}}
{e^{-y_{i}X_{i}|_{k}}} = -\sum_{k}p_{k}\ln p_{k}
\end{eqnarray}
which is the Gibbs-Shannon entropy. It has been used by many authors, since the pioneering
work by Jaynes \cite{Jaynes}, to develop ensemble theory.

\subsection{Theory of Fluctuations}
\label{fluctu}

In this section we follow an analogous treatment to the proposed by M\"unster to develop
fluctuation theory \cite{Munster}. We consider macroscopic systems in this section
(and, hence, $\open=0$).
Eq.(\ref{px}) allows to calculate the probability of having a concrete value
for a fluctuating set of macroscopic parameters $\{\alpha\}$.
To simplify notation we define the vector  $\vec{\alpha} \equiv (\alpha_{1}, \alpha_{2},...)$
containing the fluctuating variables. We have, from Eq.(\ref{px})
\begin{equation}
\mathcal{P}_{\vec{\alpha}}=\mathcal{P}_{\vec{\alpha}_{0}}e^{\ekx_{\vec{\alpha}_{0}}-\ekx_{\vec{\alpha}}}
\label{pxflu}
\end{equation}
Where $\vec{\alpha}_{0}$ is the most probable
value of the variable $\vec{\alpha}$ at an equilibrium state.
We take $\vec{\alpha}_{0}=0$ for simplicity (but with no loss of generality).
When fluctuations are small,
a Taylor expansion of $\ekx_{\vec{\alpha}}$ in Eq.(\ref{pxflu}) to second order yields
\begin{equation}
\mathcal{P}_{\vec{\alpha}} = C\mathcal{P}_{0}e^{-\frac{1}{2}
\left(\frac{\partial^{2}\ekx_{\vec{\alpha}}}{\partial \alpha_{k} \alpha_{l}}\right)_{0}
\alpha_{k}\alpha_{l}}
\label{Einstein}
\end{equation}
where $C$ is a normalization constant and subindex 0 denotes now the final equilibrium state.
Eq.(\ref{Einstein}) is the so-called Einstein fluctuation formula. It can be applied to calculate
the variances of fluctuating macroscopic parameters. In Eq.(\ref{pxflu}) $\Delta \ekx \equiv
-\left[\ekx_{\vec{\alpha}}-\ekx(0)\right]$ is to be considered as the
(dimensionless) work $W$ developed
by a subsystem on its surroundings
in evolving to an equilibrium state \cite{LandLif, Lavenda}.
This work is developed by keeping intensive environment variables constant, and, hence,
from Eq.(\ref{dgenpot})
\begin{equation}
\Delta\ekx=\delta W=-\vec{\lambda}\Delta\vec{\alpha} \label{work}
\end{equation}
where $\lambda$ must be considered as the vector of parameters conjugate to the
vector $\vec{\alpha}$.
Note that Eq.(\ref{work}) makes reference to a reversible process,
in fact $W$ can also be pictured as the minimum work $W_{min}$  which is necessary to do
on the subsystem for driving it from the equilibrium state to the nonequilibrium one. As this work
is the minimum one it is developed by following a reversible path \cite{Kubo}. It is useful to define
the matrix $\mathbf{G}$ as
\begin{equation}
G_{kl}\equiv \left(\frac{\partial^{2}\ekx_{\vec{\alpha}}}{\partial \alpha_{k} \partial \alpha_{l}}\right)_{0}=
-\left(\frac{\partial^{2}\ekx^{+}_{\vec{\lambda}}}{\partial \lambda_{k} \partial \lambda_{l}}\right)_{0}^{-1} \label{G}
\end{equation}
where in the last equality $\ekx^{+}$ is defined through a Legendre transformation of $\ekx$
by changing variables $\alpha_{k}$ (or $\alpha_{l}$) to $\lambda_{k}$ (or $\lambda_{l}$) in Eq.(\ref{work})
\begin{equation}
\Delta\ekx^{+}=-\vec{\lambda}^{*}d\vec{\alpha}^{*}+\alpha_{k}d\lambda_{k} \label{work2}
\end{equation}
The asterisk in Eq.(\ref{work2}) denotes vectors with vanishing $k$ position.
From the above considerations, Eq.(\ref{Einstein}) can be rewritten in the following way
\begin{equation}
\mathcal{P}_{\vec{\alpha}} = C\mathcal{P}_{0}e^{W}=C\mathcal{P}_{0}e^{-\frac{1}{2}
\vec{\alpha}^{T}\mathbf{G}\vec{\alpha}}
\label{Einstein2}
\end{equation}
and, therefore, from Eq.(\ref{work}) and Eq.(\ref{Einstein2})
\begin{equation}
\vec{\lambda}=\mathbf{G}\vec{\alpha}=-\frac{\partial \ln \mathcal{P}_{\vec{\alpha}}}
{\partial \vec{\alpha}} \label{lamb}
\end{equation}
For the average of the dyadic product $\vec{\alpha}\vec{\lambda}$
(in components $(\vec{\alpha}\vec{\lambda})_{kl}=\alpha_{k}\lambda_{l})$
\begin{equation}
<\vec{\alpha}\vec{\lambda}>=\int \vec{\alpha}\vec{\lambda}\mathcal{P}_{\vec{\alpha}} d\vec{\alpha}
=-\int \vec{\alpha}\frac{\partial \mathcal{P}_{\vec{\alpha}}}
{\partial \vec{\alpha}}d\vec{\alpha}
\end{equation}
and by means of integration by parts and and considering the normalization condition for the
probability distribution, we obtain
\begin{equation}
<\vec{\alpha}\vec{\lambda}>=\mathbf{U} \label{dyadic}
\end{equation}
whence
\begin{eqnarray}
<\vec{\lambda}\vec{\lambda}>&=&\mathbf{G} \label{dyadic2} \\
<\vec{\alpha}\vec{\alpha}>&=&\mathbf{G}^{-1} \label{dyadic3}
\end{eqnarray}
The following important equality holds
\begin{equation}
<\vec{\lambda}\vec{\lambda}><\vec{\alpha}\vec{\alpha}>=1 \label{Heisen}
\end{equation}
If $\alpha$ is chosen to represent 
the extensive variable $X_{s}$, we have,
from Eq.(\ref{dyadic3}) and by using Eq.(\ref{dgenpot})
\begin{equation}
<\Delta X_{s}^{2}>=\left(\frac{\partial^{2}\ekx_{X_{s}}}{\partial X_{s}^{2}}\right)_{0}^{-1}=
-\left(\frac{\partial X_{s}}{\partial y_{s}}\right)_{0}
\label{fluctuex}
\end{equation}
for the mean square fluctuation. If the conjugate intensive variable is now considered
we obtain, from Eqs. (\ref{dyadic2}) and (\ref{dgenpot}) the following result
\begin{equation}
<\Delta y_{s}^{2}>=-\left(\frac{\partial X_{s}}{\partial y_{s}}\right)_{0}^{-1}
=-\left(\frac{\partial^{2}\ekx_{y_{s}}}{\partial y_{s}^{2}}\right)_{0}^{-1}
\label{fluctuin}
\end{equation}
where superindex $+$ has been dropped in the last equality for $\ekx$ which must be understood
as the concrete specification of this function for the set of variables chosen (those
involved in the derivation plus those held constant).

The Eq.(\ref{Heisen}) for the extensive-intensive variables pair reads
\begin{equation}
<\Delta X_{s}^{2}><\Delta y_{s}^{2}>=1
\end{equation}
when the variances are evaluated by keeping constant the same environment variables.
If there are several extensive variables that can fluctuate the covariances among them
can be easily calculated from Eq.(\ref{Einstein}).
The result for the mean square covariances of two variables $X_{s_{1}}$, $X_{s_{2}}$ is
\begin{equation}
<\Delta X_{s_{1}}\Delta X_{s_{2}}>
=\left(\frac{\partial^{2}\ekx_{X_{s_{1}},X_{s_{2}}}}{\partial X_{s_{1}} \partial X_{s_{2}} }\right)_{0}^{-1}
\label{correlac}
\end{equation}

\subsection{Expressions for the customary statistical ensembles.}
\label{Ex}

In this section we apply the main expressions obtained previously
to the customary statistical ensembles.
It is useful to abbreviate notation to define the following quantities \cite{Vives}:
$\beta \equiv \frac{1}{kT}$, $\pi \equiv \frac{p}{kT}$, $\nu \equiv -\frac{\mu}{kT}$.
The following expressions were also obtained in the previous formulation by Vives and Planes
\cite{Vives}.

\subsubsection{Microcanonical Ensemble}

For this ensemble, $\{X_{j}\}=\{E,V,N\}$, $\{y_{j}\}=\{\beta,\pi,\nu\}$,
$\{X_{i}\}=\{\emptyset\}$ and $\{y_{i}\}=\{\emptyset\}$. For this system
Eqs.(\ref{genpar}) and (\ref{genent}) take the form
\begin{equation}
\ekx=-\frac{S}{k}=-\ln \Omega
\end{equation}
The values for non-environment variables can be calculated from Eq.(\ref{dgenpot})
\begin{eqnarray}
\beta &=&-\left(\frac{\partial \ekx}{\partial E}\right)_{V,N} \\
\pi &=&-\left(\frac{\partial \ekx}{\partial V}\right)_{E,N} \\
\nu &=&-\left(\frac{\partial \ekx}{\partial N}\right)_{E,V}
\end{eqnarray}
The probability distribution for a configuration $p_{k}$ is given by Eq.(\ref{probb}).
\begin{equation}
p_{k}=\frac{1}{\Omega}
\end{equation}
The variances of non-environment variables are given by Eqs.(\ref{fluctuex}) and (\ref{fluctuin})
\begin{eqnarray}
\left<\Delta \beta^{2}\right>_{V,N}&=&-\left(\frac{\partial^{2} \ekx}{\partial \beta^{2}}\right)_{V,N}^{-1}
=-\left(\frac{\partial \beta}{\partial E}\right)_{V,N}\\
\left<\Delta \pi^{2}\right>_{E,N}&=&-\left(\frac{\partial^{2} \ekx}{\partial \pi^{2}}\right)_{E,N}^{-1}
=-\left(\frac{\partial \pi}{\partial V}\right)_{E,N}\\
\left<\Delta \nu^{2}\right>_{E,V}&=&-\left(\frac{\partial^{2} \ekx}{\partial \nu^{2}}\right)_{E,V}^{-1}
=-\left(\frac{\partial \nu}{\partial N}\right)_{E,V}
\end{eqnarray}
Note, that on evaluating the fluctuations, the independent variables considered are
those held constant and the one involved in the derivation. $\ekx$ is considered
as a function of these independent variables and hence, Eq.(\ref{dgenpot}) is used each time
with different sets of independent variables. For example, on evaluating the variance of $\beta$,
variables $\beta$, $V$, $N$ are taken as independent in writing Eq.(\ref{dgenpot}), but, on evaluating
the variance of $\pi$, the independent variables are $\pi$, $E$, $N$. This means that $\ekx$ is different
in each case, but if one uses the general definition Eq.(\ref{dgenpot}) for $\ekx$, these differences
are of no formal interest as a complete set of environment variables specify those which are not natural
in the system, through Eq.(\ref{dgenpot}), in general.

\subsubsection{Canonical Ensemble}

In this case $\{X_{j}\}=\{V,N\}$, $\{y_{j}\}=\{\pi,\nu\}$,
$\{X_{i}\}=\{E\}$ and $\{y_{i}\}=\{\beta \}$. For this system
Eqs.(\ref{genpar}) and (\ref{genent}) take the form
\begin{equation}
\ekx=\beta F=\beta(<E>-TS)=-\ln Z
\end{equation}
where $F$ is the Helmholtz free energy and $Z \equiv g(\beta, N, V)$
is the partition function of the system.

The values for non-environment variables can be calculated again from Eq.(\ref{dgenpot})
\begin{eqnarray}
<E> &=&\left(\frac{\partial \ekx}{\partial \beta}\right)_{V,N} \\
\pi &=&-\left(\frac{\partial \ekx}{\partial V}\right)_{\beta,N} \\
\nu &=&-\left(\frac{\partial \ekx}{\partial N}\right)_{\beta,V}
\end{eqnarray}
The probability distribution for a configuration $p_{k, E}$ in which the system has energy $E$,
is given by Eq.(\ref{probb}) which in this case takes the form
\begin{equation}
p_{k,E}=\frac{e^{-\beta E}}{Z}
\end{equation}
The variances of non-environment variables are given by Eqs.(\ref{fluctuex}) and (\ref{fluctuin})
\begin{eqnarray}
\left<\Delta E^{2}\right>_{V,N}&=&\left(\frac{\partial^{2} \ekx}{\partial E^{2}}\right)_{V,N}^{-1}
=-\left(\frac{\partial E}{\partial \beta}\right)_{V,N}\\
\left<\Delta \pi^{2}\right>_{\beta,N}&=&-\left(\frac{\partial^{2} \ekx}{\partial \pi^{2}}\right)_{\beta,N}^{-1}
=-\left(\frac{\partial \pi}{\partial V}\right)_{\beta,N}\\
\left<\Delta \nu^{2}\right>_{\beta,V}&=&-\left(\frac{\partial^{2} \ekx}{\partial \nu^{2}}\right)_{E,V}^{-1}
=-\left(\frac{\partial \nu}{\partial N}\right)_{\beta,V}
\end{eqnarray}

\subsubsection{Grand Canonical Ensemble}

For a system that can exchange energy and particles with a heat bath at $T$ and $\mu$, we have
$\{X_{j}\}=\{V\}$, $\{y_{j}\}=\{\pi\}$,
$\{X_{i}\}=\{E,N\}$ and $\{y_{i}\}=\{\beta,\nu \}$. For this system
Eqs.(\ref{genpar}) and (\ref{genent}) take the form
\begin{equation}
\ekx=\beta(<E>-TS)+\nu <N>=-\ln \Xi
\end{equation}
where $\Xi \equiv g(\beta, \nu, V)$
is the grand partition function of the system.

The values for non-environment variables can be calculated again from Eq.(\ref{dgenpot})
\begin{eqnarray}
<E> &=&\left(\frac{\partial \ekx}{\partial \beta}\right)_{V,\nu} \\
\pi &=&-\left(\frac{\partial \ekx}{\partial V}\right)_{\beta,\nu} \\
<N> &=&\left(\frac{\partial \ekx}{\partial \nu}\right)_{\beta,V}
\end{eqnarray}
The probability distribution for a configuration $p_{k, E, N}$ in which the system has energy $E$
and number of particles $N$ is, from Eq.(\ref{probb})
\begin{equation}
p_{k,E,N}=\frac{e^{-\beta E-\nu N}}{\Xi}
\end{equation}
The variances of non-environment variables are given by Eqs.(\ref{fluctuex}) and (\ref{fluctuin})
\begin{eqnarray}
\left<\Delta E^{2}\right>_{V,\nu}&=&\left(\frac{\partial^{2} \ekx}{\partial E^{2}}\right)_{V,\nu}^{-1}
=-\left(\frac{\partial E}{\partial \beta}\right)_{V,\nu}\\
\left<\Delta \pi^{2}\right>_{\beta,\nu}&=&-\left(\frac{\partial^{2} \ekx}{\partial \pi^{2}}\right)_{\beta,\nu}^{-1}
=-\left(\frac{\partial \pi}{\partial V}\right)_{\beta,\nu}\\
\left<\Delta N^{2}\right>_{\beta,V}&=&\left(\frac{\partial^{2} \ekx}{\partial N^{2}}\right)_{\beta,V}^{-1}
=-\left(\frac{\partial N}{\partial \nu}\right)_{\beta,V}
\end{eqnarray}

Since for this ensemble there are two extensive fluctuating variables, we can calculate also the correlation
between them, from Eq.(\ref{correlac})
\begin{eqnarray}
<\Delta E \Delta N>
&=&\left(\frac{\partial^{2}\ekx}{\partial E \partial N}\right)_{V}^{-1} \nonumber \\
&=&-\left(\frac{\partial N}{\partial \beta}\right)_{V,\nu}
=-\left(\frac{\partial E}{\partial \nu}\right)_{\beta,V}
\end{eqnarray}

\section{Squeezed Statistical Mechanics: A generalization of thermostatistics
to systems out of BG equilibrium.}
\label{generalization}

Hitherto we have considered systems with no capability to choose between the attainable configurations
compatible with the constraints. The systems treated here fill their attainable configurations depending
on their degree of attainability, as specified by the physical constraints imposed to the system.
The features of the constraints have been the only criterion for the system to decide what configurations
adopt. As we have seen these constraints can be extensive or intensive, and are imposed to the system
thus restricting its behavior.

There are complex systems in nature that exhibits self-organization and have the ability of taking
decisions. These systems can restrict (or enlarge) the total number of configurations at their disposal
by selecting only a group of them or expanding their spectrum against constraints.
An analogous case can be that of systems in nonequilibrium situations for which
the total number of configurations of the phase space are not necessarily available.

We thus see that new information is required in order to account for these systems.
A possible way to introduce this information is to modify the definition of $\ekx$
by introducing
\emph{not the complete set of configurations but the actual one} $\widetilde{g}=h_{q}(g)$.
The function $h$ specifies
the information of the actual set of attainable configurations for the system
in reference to those attainable from the only specification of the constraints ($g$)
and then act as an interface between equilibrium and complex systems
or nonequilibrium situations.
The parameter $q$ is introduced for the sake of generality
to classify different statistics under
the common denominator of a superstatistics, specified by the squeezing function $h$.
We remark that the function $h$ can have a more complex dependence on system variables
and situations. We consider here, however, the case of one-parameter dependent squeezing functions, in which the complexity is accounted for through an entropic
parameter $q$ and the specification of the function $h$. Obvious extensions are considering 
$n$-parameter dependent squeezing functions $h_{q_{1},... q_{n}}$. A theory for these can be developed following analogous lines to those we discuss below.
The convenience of introducing at least one additional entropic parameter stems in the diversity
of the dynamics of complex systems. We introduce the function $h_{q}$ as an additional postulate

\begin{postulate}
\label{p3}
The actual classes of configurations $\widetilde{g}$ 
of a complex system can be determined by a convenient positive squeezing function $h_{q}(g)$ of
the BG classes $g$ by means of the relation $\widetilde{g}=h_{q}(g)$. 
This function has a well defined inverse function $H_{q}$, being $h_{q}(H_{q}(x))=H_{q}(h_{q}(x))=x.$
\end{postulate}

The set of fundamental equations (\ref{genpar}), (\ref{dgenpot}) 
and (\ref{genent}) can be rewritten then as
\begin{eqnarray}
\ekx &=&-\ln \widetilde{g}_{\{X_{j}\} \cup \{y_{i}\}}=-\ln h_{q}(g_{\{X_{j}\} \cup \{y_{i}\}}) \label{qgenpar} \\
d\ekx &=&-y_{j,obs}dX_{j}+ X_{i,obs}dy_{i} \label{qdgenpot} \\
\J &=& -\ekx+y_{i}X_{i,obs}=
-\ekx+y_{i}\left(\frac{\partial \ekx}{\partial y_{i}} \right)_{\{X_{j}\},\{y_{i^{*}}\}} \nonumber \\
&=& \ln \widetilde{g}_{\{X_{j}\}}= \ln h_{q}(g_{\{X_{j}\}}) \label{qgenent} \\
\open &=&
-\ekx-y_{j,obs}X_{j}=
-\ekx+X_{j}\left(\frac{\partial \ekx}{\partial X_{j}} \right)_{\{X_{j^{*}}\},\{y_{i}\}} \nonumber\\
&=& \ln \widetilde{g}_{\{y_{i}\}}=\ln h_{q}(g_{\{y_{i}\}})
\label{qgenopen}\\
\widetilde{g}&=&h_{q}(g) \label{genrest}
\end{eqnarray}
If we compare these expressions with
Eqs.(\ref{genpar}), (\ref{dgenpot}), (\ref{genent}) and (\ref{Hillgenent}),
we see that the modification introduced concerns only the attainable configurations
by the system through Eq. (\ref{genrest}) and anything else. Additive decomposition of the actual
(squeezed) classes 
(as the fulfilled by the Boltzmann-Gibbs equilibrium classes) does no longer
hold in principle, however. Expressions like Eqs.(\ref{mic0}) and (\ref{mic}) cannot be written
for the ``squeezed" classes $\widetilde{g}$. This fact has its deep grounds in the underlying 
nonlinear kinetics exhibited by these systems (see below).

We call the framework described by Eqs. (\ref{qgenpar}) to (\ref{genrest}) Squeezed
Statistical Mechanics. The term ``squeezed" makes reference to the 
function $h_{q}$ that deforms the Boltzmann-Gibbs equilibrium phase space.
The expressions above are valid for any statistical system and can be
considered as the skeleton of the thermostatistics presented here. Nonextensive thermodynamics demands
$\open \ne 0$ but it does not necessarily imply a nonextensive statistics.
Nonextensive statistics is obtained when the function $h_{q}$ is different to the identity.
In the case of extensive statistics Eqs. (\ref{qgenpar}) to (\ref{genrest}) are equal
to Eqs.(\ref{genpar}),(\ref{dgenpot}), (\ref{genent}), and (\ref{Hillgenent}) with
$h_{q}(x)=x$. The main motivation to introduce $h_{q}(x) \ne x$
is a practical one: it is usually more feasible
counting over the complete distribution of configurations (i.e. the equilibrium Boltzmann closure)
than over the incomplete distribution which constitutes the nonequilibrium closure. When this is done,
Boltzmann's principle cannot be applied with the equilibrium closure alone and \emph{an additional
information is required to relate the actual total number of configurations to that
of the equilibrium closure}. \emph{The different statistics arise
when one uses the equilibrium (Boltzmann) closure to establish probability distributions for
complex systems or nonequilibrium situations. As a result of that, the equilibrium closure
and classes of configurations are squeezed and rearranged, and no longer resemble those of BG equilibrium}.
We show this point below.

Gibbs fundamental equation written in the generalized form, Eq. (\ref{qdgenpot}), is
formally identical to the obtained above for the ordinary thermostatistics, Eq.(\ref{dgenpot}).
The robustness of the Legendre transform structure and Gibbs fundamental equation for arbitrary
entropic forms and mean values has been recently pointed out in the references \cite{Yamano}
in the framework of the Jaynes approach.

The actual total number of configurations $\widetilde{g}$ is then to be understood
as dependent on the total number of configurations $g$ compatible with the constraints and,
maybe, on new parameters. We emphasize the fact that configurations
$\widetilde{g}$ represent, in any case, the \emph{actual}
configurations of the system and \emph{not} those resulting from counting over the complete
distribution $g$ which are related to a system under the hypothesis of the molecular chaos (i.e. with no organization and no ability of taking decisions)
at equilibrium. This case has been considered in previous sections. Again, the methods used in previous sections to formulate
equilibrium statistical mechanics can be identically translated here to the extension to nonequilibrium
situations and complex systems.

\subsection{Generalized ensemble theory.}

Let us write the squeezing function in terms of the actual and the BG characteristic classes
\begin{equation}
\widetilde{g}_{\{X_{j}\} \cup \{y_{i}\}}=h_{q}\left(g_{\{X_{j}\} \cup \{y_{i}\}}\right)
\end{equation}
The function $h_{q}$ is equal to the identity for BG statistics. We have, from Eq.(\ref{qgenpar})
\begin{eqnarray}
\ekx&=&-\ln h_{q}\left(g_{\{X_{j}\} \cup \{y_{i}\}}\right) \\
\ekx_{\{X_{i}\}}&=&-\ln h_{q}\left(g_{\{X_{j}\} \cup \{y_{i}\} \cup \{X_{i}\}}\right) \label{superekx}
\end{eqnarray}
and thus,
if Eq.(\ref{qgenent}) is used
\begin{eqnarray}
\ekx _{\{X_{i}\}}&=&y_{i}X_{i}-\J _{\{X_{i}\}}
=y_{i}X_{i}-\ln h_{q} \left(g_{\{X_{j}\} \cup \{X_{i}\}}\right)\nonumber \\
&=&-\ln h_{q} \left(g_{\{X_{j}\} \cup \{X_{i}\}}\right)e^{-y_{i}X_{i}}
\label{superostia}
\end{eqnarray}
hence, from Eqs.(\ref{superekx}) and (\ref{superostia})
\begin{equation}
h_{q}\left(g_{\{X_{j}\} \cup \{y_{i}\} \cup \{X_{i}\}}\right)=
h_{q}\left(g_{\{X_{j}\} \cup \{X_{i}\}}\right)e^{-y_{i}X_{i}} \label{superend}
\end{equation}
Let us consider the inverse function of $h_{q}$, $H_{q}$, so that $H_{q}(h_{q}(x))=x$.
We have
\begin{equation}
g_{\{X_{j}\} \cup \{y_{i}\} \cup \{X_{i}\}}=H_{q}\left(h_{q}\left(g_{\{X_{j}\} \cup \{X_{i}\}}\right)e^{-y_{i}X_{i}}\right) \label{pf}
\end{equation}
and, therefore,
\begin{equation}
\ekx=-\ln h_{q}\left[\sum_{\{X_{i}\}}H_{q}\left(h_{q}\left(g_{\{X_{j}\} \cup \{X_{i}\}}\right)e^{-y_{i}X_{i}}\right)\right]
\label{ekkx}
\end{equation}
This is the central equation of generalized ensemble theory. It has to be supplemented to the following for the two entropic
forms that we have been considering $\J$ and $\open$, the last being of relevance for small systems. 
\begin{eqnarray}
\J&=& \ln h_{q}(g_{\{X_{j}\}})=\ln h_{q}\left[\sum_{\{X_{i}\}}g_{\{X_{j}\} \cup \{X_{i}\}}\right] 
\label{enn} \\
\open&=& \ln h_{q}(g_{\{y_{i}\}})=\ln h_{q}\left[\sum_{\{X_{i}\}}H_{q}\left(h_{q}\left(g_{\{X_{i}\}}\right)e^{-y_{i}X_{i}}\right)\right]
\label{opp}
\end{eqnarray}
All these expressions equal to those of equilibrium when the squeezing function is equal to the identity.

\subsection{Extensive and nonextensive representations. Subdivision property}

The above results suggest that there exist two convenient 
representations when nonequilibrium or complex systems are considered. These two representations are equal only
for systems following BG statistics. We have, on one hand, the extensive representation specified by the actual classes
$\widetilde{g}$. In terms of these the entropy, specified by Eq. (\ref{qgenent}), is extensive
\begin{equation}
\J_{A+B}=\ln \widetilde{g}_{A+B}= \ln \widetilde{g}_{A}\widetilde{g}_{B}=\ln \widetilde{g}_{A}+\ln \widetilde{g}_{B}=\J_{A}+\J_{B}
\end{equation}
From this extensivity property we have another very interesting property. The subdivision of the actual classes 
of a composite system in BG classes of independent subsystems has the following form
\begin{equation}
\widetilde{g}_{A+B}= h_{q}(g_{A})h_{q}(g_{B})
\end{equation}
We call this property of factorization an actual class in BG subclasses \emph{subdivision property} of the actual class.

We have also a representation specified by the BG classes which leads to nonextensivity in the entropic form
(as BG classes are not the proper representation of the system). This can be clearly seen from Eq. (\ref{qgenent}). If 
the inverse function $H_{q}$ is used one straightforwardly obtains
\begin{equation}
\J_{A+B}=\ln h_{q}(g_{A}g_{B})=\ln h_{q}\left[H_{q}(e^{\J_{A}})H_{q}(e^{\J_{B}})\right] \ne \J_{A}+\J_{B} \label{nextp}
\end{equation}
Although the BG classes are not the proper classes of the system, it is very convenient working with them due to the
very useful properties these have and which are lacking in general in the proper actual classes, the property
of linear decomposability Eq.(\ref{mic}) being an example. Furthermore, the equilibrium classes can be known better as they belong to the
complete description of the BG closure.

\subsection{Probability}

We maintain unchanged the definition of probability we given at equilibrium (see section \ref{prob}). 
The reason is that we usually use the BG classes to evaluate probabilities since these constitute
the complete closure of the phase space of the system. The form of the probability distributions is 
no longer that of BG thermostastics, however: the classes are rearranged as we show below giving rise
to bunches of distributions that can be classified within a concrete superstatistics. Each of the distributions
has to be understood as a different way of partitioning the same equilibrium closure. For the probability
we have, then
\begin{eqnarray}
&&\mathcal{P}_{\{X_{s}\}}=\frac{g_{\{X_{j}\} \cup \{y_{i}\} \cup \{X_{s}\}}}
{g _{\{X_{j}\}\cup \{y_{i}\}}}=
\frac{g_{\{X_{j}\} \cup \{y_{i}\} \cup \{X_{s}\}}}
{\sum_{\{X_{s}\}}g_{\{X_{j}\} \cup \{y_{i}\} \cup \{X_{s}\}}} \nonumber \\
&=&\frac{H_{q}\left(h_{q}\left(g_{\{X_{j}\} \cup \{X_{s}\}}\right)e^{-y_{s}X_{s}}\right)}
{\sum_{\{X_{s}\}}H_{q}\left(h_{q}\left(g_{\{X_{j}\} \cup \{X_{s}\}}\right)e^{-y_{s}X_{s}}\right)} 
\label{pflug}\\  
&& \nonumber \\
&&p_{k,\{X_{s}\}}=\frac{\mathcal{P}_{\{X_{s}\}}}{g _{\{X_{j}\}\cup \{X_{s}\}}} 
=\frac{\frac{H_{q}\left(h_{q}\left(g_{\{X_{j}\} \cup \{X_{s}|k\}}\right)e^{-y_{s}X_{s}|k}\right)}{g _{\{X_{j}\}\cup \{X_{s}|k\}}}}
{\sum_{k}\frac{H_{q}\left(h_{q}\left(g_{\{X_{j}\} \cup \{X_{s}|k\}}\right)e^{-y_{s}X_{s|k}}\right)}{g _{\{X_{j}\}\cup \{X_{s}|k\}}}}
\label{pbgen}
\end{eqnarray}
where $X_{s}|_{k}$ denote the value of variable $X_{s}$ in the configuration $k$
and $g_{\{X_{j}\} \cup \{X_{s}|_{k}\}}$ is the total number of microcanonical
configurations within the class to which the configuration $k$ belongs. 
The generalized Boltzmann factor $B_{\{X_{i}\}}$ can then be defined as
\begin{equation}
B_{\{X_{i}\}}\equiv \frac{g_{\{X_{j}\} \cup \{y_{i}\} \cup \{X_{i}\}}}
{g_{\{X_{j}\} \cup \{X_{i}\}}}=\frac{H_{q}\left(h_{q}\left(g_{\{X_{j}\} \cup \{X_{i}\}}\right)
e^{-y_{i}X_{i}}\right)}{g_{\{X_{j}\} \cup \{X_{i}\}}} \label{superB}
\end{equation}
From Eqs.(\ref{superostia}), (\ref{ekkx}) and (\ref{enn}) we obtain
\begin{eqnarray}
\mathcal{P}_{\{X_{s}\}}&=&\frac{H_{q}(e^{-\ekx_{\{X_{s}\}}})}{H_{q}(e^{-\ekx})}
\label{genfl}\\
p_{k,\{X_{s}\}}&=&\frac{H_{q}(e^{-\ekx_{\{X_{s}|k\}}})}{H_{q}(e^{\J_{\{X_{s}|k\}}})H_{q}(e^{-\ekx})}
\label{genfl2}
\end{eqnarray}
These probability distributions are normalized.

For the observed values of extensive non-environment variables, we have from Eqs.(\ref{qdgenpot}),
(\ref{ekkx}) and (\ref{pbgen}) and after a straightforward calculation that
\begin{eqnarray}
X_{i,obs}&=&
\left(\frac{\partial \ekx}{\partial y_{i}} \right)_{\{X_{j}\},\{y_{i^{*}}\}}
= \sum_{\{X_{i}\}}\frac{f(g_{\{X_{j}\} \cup \{y_{i}\}})}{f(\mathcal{P}_{\{X_{i}\}}g_{\{X_{j}\} \cup \{y_{i}\}})}
X_{i} \nonumber \\
&=&\sum_{k}\frac{f(g_{\{X_{j}\} \cup \{y_{i}\}})}{g_{\{X_{j}\} \cup \{X_{i|k}\}}f(p_{k}g_{\{X_{j}\} \cup \{X_{i|k}\}}g_{\{X_{j}\} \cup \{y_{i}\}})}
X_{i|k}
\label{genobs}
\end{eqnarray}
where $f(x)=dh_{q}(x)/dx$ is the first derivative of the squeezing function. We note that the definition of the mean field
value is very different from that of BG equilibrium although the same definition for probability has been used. This is
due to the fact that the classes of configurations of the equilibrium closure have been rearranged and the partition is
different now.

\subsection{Connection with Beck-Cohen superstatistics}

A very recent approach to generalize statistical mechanics to nonequilibrium from the bottom up
is that of Beck and Cohen \cite{Beck}. The following discussion
will be referred to the canonical ensemble only, as is done in Beck-Cohen's paper.
The approach followed by the authors
is based on the definition of the generalized Boltzmann factor $B(E)$
\begin{equation}
B_{E}=\int_{0}^{\infty}d\beta'f(\beta')e^{-\beta'E} \label{superBeck}
\end{equation}
where $f(\beta')$ is a function of the thermal Lagrange parameter $\beta'$ which is a normalized
probability density that allows to build a normalizable statistics which must be reduced
to BG-statistics when there are no fluctuations in $\beta'$. In fact
the choice $f(\beta')=\delta(\beta'-\beta)$ allows to recover the ordinary Boltzmann factor
$e^{-\beta E}$. We can relate easily the formalism of Beck and Cohen to the presented here.
From Eqs.(\ref{superB}) and (\ref{superBeck}) we obtain
\begin{equation}
f(\beta')=L^{-1}_{\beta}\left[\frac{H_{q}\left(h_{q}\left(g_{\{N,V\} \cup \{E \}}\right)
e^{-\beta E}\right)}{g_{\{N,V\} \cup \{E\}}}\right]
\end{equation}
where $L^{-1}_{\beta}$ denotes inverse Laplace transform with respect to variable $\beta$.
We see that an interesting property (but not, in fact, a requirement) for the function $H$ is to have a well
defined inverse Laplace transform. 

\subsection{Zeroth law of thermodynamics: a pathway to the relevant statistics.}

Within the extensive framework, if two systems I and II are put in contact attaining thermodynamical equilibrium their 
Lagrange parameters become equal
\begin{equation}
y_{j}^{I}=y_{j}^{II} \label{text}
\end{equation}
From Eqs.(\ref{qgenpar}) and (\ref{qdgenpot}), this equality can be rewritten within the nonextensive 
representation (microcanonical ensemble) as 
\begin{equation}
\frac{d\ln h_{q}(g_{X_{j}}^{I})}{d\ln g_{X_{j}}^{I}}y_{j,BG}^{I}=
\frac{d\ln h_{q}(g_{X_{j}}^{II})}{d\ln g_{X_{j}}^{II}}y_{j,BG}^{II} \Rightarrow  
f(g_{X_{j}}^{I})y_{j,BG}^{I}=f(g_{X_{j}}^{II})y_{j,BG}^{II} \label{tnext}
\end{equation}
where $y_{j,BG}=-d\ln g_{\{X_{j}\}}/dX_{j}$ is a BG Lagrange parameter which, if the thermal one is considered, 
can be measured by an ordinary thermometer following BG thermostatistics. We arrive to the following conclusion: the temperature measured
by means of a BG thermometer for systems following the generalized zeroth law principle is, in general, \emph{different}
for each of the systems in contact if those are of different nature (i.e. being described by different statistics).
This is not surprising since temperature measured by a thermometer depends on calibration, having a statistical origin,
and we are describing non conventional situations \emph{out of BG equilibrium but following a more general definition
of thermal equilibrium in which only temperatures measured by proper thermometers are equal}.
If a system does not follow BG statistics, a proper thermometer should be developed to measure the \emph{actual} 
temperature at equilibrium. Stated in other way, for two systems in (general) thermal equilibrium, the proper thermometer of each
system measures the same temperature but \emph{temperatures measured by BG thermometers are different (in general) 
if those are of different statistical nature}.

Let us consider now that the system denoted by II is described 
by BG statistics $(f(g_{X_{j}}^{II})=1)$ and 
it is in equilibrium with system I which is a complex one. Eq. (\ref{tnext}) can be integrated to give
\begin{equation}
h_{q}(g_{X_{j}}^{I})=h_{q}(g_{X_{j}}^{I*})
e^{\int_{\ln g_{X_{j}}^{I*}}^{\ln g_{X_{j}}^{I}}\frac{y_{j,BG}^{I}}{y_{j,BG}^{II}}d \ln g_{X_{j}}^{'I}}
\label{mech}
\end{equation}
where $h_{q}(g_{X_{j}}^{I*})$ is a reference state. If we take $h_{q}(1)=1$ when there is only
one attainable microstate, having no squeezing in this case (which seems reasonable as all superstatistics should then 
collapse) and we take series of measurements of $y_{j,BG}$ for both systems in (generalized) thermal equilibrium  
calculating their ratio as a function of the number of BG attainable microstates of the complex system, Eq.(\ref{mech})
provides a mechanism to determine \emph{a priori} the statistics that the latter follows.

\subsection{Irreversibility and generalized Boltzmann equation}

The $actual$ attainable classes of microstates $\tilde{g}$ introduced here can be directly related to the underlying
nonlinear kinetics in the squeezed statistics under consideration \cite{Kan1}. 
The kinetic interaction principle first stated by Kaniadakis \cite{Kan2} leads to a generalization of the
Boltzmann's kinetic equation allowing to consider more complex systems. Our formalism is completely consistent with
the previous Kaniadakis theory, which in fact, sheds some light on the meaning of the squeezing function $h_{q}$. 

The number of particles in a volume $d\vec{r}d\vec{v}$ (where $\vec{r}$ and $\vec{v}$ 
are the position and the velocity of a given position inside a fluid) is given, at time $t$ 
in terms of the one-body density function $F(\vec{r}, \vec{v}, t)$, by the product 
$F(\vec{r}, \vec{v}, t)d\vec{r}d\vec{v}$. If we use the above established subdivision property 
for the density $F$ we can assume that the number of pairs of particles in the configuration 
space element $d\vec{r}$ and velocities $d\vec{v}_{1}$ and
$d\vec{v}_{2}$ is given by the following relation
\begin{equation}
h_{q}[F(\vec{r}, \vec{v_{1}}, t)]h_{q}[F(\vec{r}, \vec{v_{2}}, t)]d\vec{v}_{1}d\vec{v}_{2}
d\vec{r}d\vec{r}
\end{equation}
This assumption is reasonable if we realize that $\vec{v}$ specifies a general \emph{actual} microstate of one 
system and we can thought this system to be subdivided in parts at different microstates
\emph{counted in the BG closure} with velocities $\vec{v}_{1}$ and $\vec{v}_{2}$.
In the case of BG systems (where $h_{q}$ is the identity) the assumption coincides 
with the celebrated \emph{Stosszahlansatz} (molecular
chaos hypothesis) due to Boltzmann. It cannot be justified from first principles and it is certainly
nor necessary nor true for all time \cite{Wehrl}. We are considering here a more general form for the number 
of pairs of particles arising from squeezing of the different elements of phase space and leading to a general
factorization of the one particle distribution function. It also leads to a very general
nonlinear interaction which has been discussed previously \cite{Kan2} and from which the following 
generalized Boltzmann equation is obtained (if we replace $\kappa(f)$ in the discussion of \cite{Kan2} by $h_{q}(F)$)
\begin{eqnarray}
&&\left(\frac{\partial}{\partial t}+\vec{v}\frac{\partial}{\partial \vec{x}}-\frac{1}{m}
\frac{\partial V(\vec{x})}{\partial \vec{x}}
\frac{\partial}{\partial \vec{v}}\right)F= \int_{\Re}d^{n}v'd^{n}v_{1}d^{n}v'_{1}\Big\{\left[h_{q}(F')h_{q}(F'_{1})-h_{q}(F)h_{q}(F_{1})\right]\times 
\nonumber \\
&&
\left.\xi(F',F)\xi(F'_{1},F_{1})
T(t,\vec{x},\vec{v},\vec{v'},\vec{v_{1}},\vec{v'_{1}})\right\}
\end{eqnarray}
where $V(\vec{x})$ is an external potential, $\vec{x}$ is the position vector of a site in the fluid and, 
$\vec{v},\vec{v'},\vec{v_{1}},\vec{v'_{1}}$ are the initial and final velocities of particles on this site 
and on another one being both in interaction. $F$, $F'$, $F_{1}$ and $F'_{1}$ are the initial and final particle
densities of each site. $T(t,\vec{x},\vec{v},\vec{v'},\vec{v_{1}},\vec{v'_{1}})$ is a factor which depends
only on the nature of the two body particle interaction (proportional to the cross section) and 
$\xi(F,F')=\xi(F',F)$ is an arbitrary symmetric function. This equation reduces to the traditional Boltzmann
equation if $h_{q}(F)=F$ and $\xi(F',F)=\xi(F'_{1},F_{1})=1$. 
\begin{equation}
\left(\frac{\partial}{\partial t}+\vec{v}\frac{\partial}{\partial \vec{x}}-\frac{1}{m}
\frac{\partial V(\vec{x})}{\partial \vec{x}}
\frac{\partial}{\partial \vec{v}}\right)F= \int_{\Re}d^{n}v'd^{n}v_{1}d^{n}v'_{1}\left[F'F'_{1}-FF_{1}\right] 
T(t,\vec{x},\vec{v},\vec{v'},\vec{v_{1}},\vec{v'_{1}})
\end{equation}
Importantly, if $h_{q}(g)$ is the identity,
the variational principle leads to the Maxwell-Boltzmann distribution regardless of the value of the function 
$\xi$. This result indicates that there is an infinity of ways (one for any choice of $\xi$) to obtain 
the Maxwell-Boltzmann distribution.  

If $\mu$ is the chemical potential of the particles and $U(\vec{v})$ their kinetic energy,
the generalized Boltzmann equation implies that the Lyapunov functional $\mathcal{H}$ given by \cite{Kan2}
\begin{equation}
\mathcal{H}=-\int_{\Re}d^{n}xd^{n}v\int dF \ln h_{q}(F)-\beta \int_{\Re}d^{n}xd^{n}v \left[V(x)+U(v)-\mu \right] F
\end{equation}
is non decreasing with time if $f(F) \equiv dh_{q}(F)/dF \ge 0$ and hence as the second integral in the r.h.s of this
equation is a conserved quantity, the entropy $S=-\int_{\Re}d^{n}xd^{n}v\int dF \ln h_{q}(F)$ 
satisfies also that $dS/dt \ge 0$. The condition $f(F)$ was advanced implicitly (for positive BG temperatures) 
when studying the generalized zeroth law (see above) provided that $h_{q}$ is a positive function in Eq. (\ref{tnext}).
We remark that being $h_{q}$ an arbitrary function, H-theorem is thus verified for a very large class of nonlinear
systems. This kinetical picture completes the above formulation of Squeezed Statistical Mechanics, in which a general
thermodynamic framework for extensive/nonextensive macroscopic/small systems has been provided.

\subsection{``Squeezing and rearranging"}

In the above sections we have formulated BG statistical mechanics through a new method which can be called "rearranging". 
Related to it are the concepts of BG classes of configurations, linear decomposability and of the
existence of a function for any system which is a monotonic decreasing function when increasing elements in the 
characteristic class. The key step in the method is that which goes from Eq. (\ref{ekx2}) to Eq.(\ref{ekx5}).

In extending BG thermostatistics to other many relevant situations we have shown that a previous step is needed in order
to define the statistics. This previous step is called "squeezing" and consists on the specification of the function $h_{q}$
and calculating (if possible) its inverse $H_{q}$ and its first derivative $f$. Then a general version of the rearranging 
step, Eqs. (\ref{superekx}) to (\ref{superend}) leads to the decomposition of the equilibrium classes and, hence, 
to a workable version of the non conventional statistics at hand.

\section{Application: Tsallis thermostatistics}
\label{app}

A recent important effort to extend statistical mechanics to complex systems has been due
to Tsallis \cite{Tsallis}. From ideas taken, in part,
from multifractal analysis and information theory,
Tsallis proposed a definition for the
entropy based on the parameter $q$ and after applying the condition of extremum for the entropy, he
provided an extended formulation of statistical mechanics.
In the following paragraphs we develop Tsallis statistics on the context of our framework
by introducing a particular choice on the q-squeezing function for the total attainable configurations compatible with the constraints. 

Let us consider a complex system (I) in (generalized) thermal equilibrium with a BG thermal bath (II) at $\beta_{BG, II}$. 
Measurements with a BG thermometer of $\beta_{BG, I}$ are taken over the complex system by varying (for example) 
its volume. Let be the BG microstates of the complex system known each time. These are all those 
allowed by the constraints, that can be controlled externally.
Let us further consider that a power law (we abbreviate by $g_{I}$ the number of attainable 
microstates of the complex system) 
controls the quotient of Lagrange parameters (which is not unreasonable due to the ubiquity of power laws in nature)
\begin{equation}
\frac{\beta_{BG, I}}{\beta_{BG, II}} \sim g_{I}^{1-q}
\end{equation}
the exponent $1-q$ is introduced here for convenience. If $q=1$ we then would have BG equilibrium between two BG systems.
If this power law is introduced in Eq. (\ref{mech}) and we integrate from $h_{q}(g_{X_{j}}^{I*})=h_{q}(1)=1$ we obtain
the following form for the squeezing function $h_{q}(x)$
\begin{equation}
h_{q}(x)=e^{\frac{{x}^{1-q}-1}{1-q}}
\end{equation}
The inverse function $H_{q}(x)$ and the first derivative $f_{q}(x)$ can be computed straightforwardly
\begin{eqnarray}
H_{q}(x)&=&[(1-q)\ln x+1]^{1/(1-q)} \label{inverse} \\
f_{q}(x)&=&x^{-q} \label{firstd}
\end{eqnarray}
This completes the "squeezing" step. 
The knowledge of these three functions is enough to develop the thermostatistical framework. 
We use them in all what follows. The following relations hold
\begin{eqnarray}
\ln(h_{q}(x))&=& \ln_{q}x \label{qlog} \\
H_{q}(e^{x})&=& e_{q}^{x} \label{qexp} 
\end{eqnarray}
In these expressions 
$\ln_{q}x$ and $e_{q}^{x}$ are, respectively the q-logarithm and the q-exponential
functions first defined by Tsallis
\begin{eqnarray}
\ln_{q}x & \equiv & \frac{x^{1-q}-1}{1-q} \\
e_{q}^{x} & \equiv & [(1-q)x+1]^{1/(1-q)} 
\end{eqnarray}

\subsection{Entropy, characteristic function and ensemble theory. Nonextensivity}

From Eqs. (\ref{ekkx}) and (\ref{enn}) we immediately obtain 
\begin{eqnarray}
\ekx&=&-
\frac{g_{\{X_{j}\} \cup \{y_{i}\}}
^{1-q}-1}{1-q} \label{qekx} \\
\J&=&
\frac{g_{\{X_{j}\}}^{1-q}-1}{1-q}
\label{qentr}
\end{eqnarray}
for the characteristic function and the entropy, respectively. The form of the characteristic class 
$g_{\{X_{j}\} \cup \{y_{i}\}}$ can be obtained easily from Eq. (\ref{pf})
\begin{equation}
g_{\{X_{j}\} \cup \{y_{i}\}}=\sum_{\{X_{i}\}}g_{\{X_{j}\} \cup \{y_{i}\} \cup \{X_{i}\}}=
\sum_{\{X_{i}\}}
g_{\{X_{j}\} \cup \{X_{i}\}}
\left(1+\frac{y_{i}X_{i}(q-1)}{g_{\{X_{j}\} \cup \{X_{i}\}}^{1-q}}
\right)^{\frac{1}{1-q}} \label{tsalgenpar}
\end{equation}

We observe that, for the microcanonical ensemble, Tsallis microcanonical partition function
coincides with the Boltzmann one ($\Omega$), as it is easily seen from Eq.(\ref{tsalgenpar})
by considering that for a system in the microcanonical ensemble $\{X_{i}\}=\{y_{i}\}=\{\emptyset \}$.

The nonextensivity property for the entropy in terms of the BG classes can be also straightforwardly
obtained from Eq. (\ref{nextp})
\begin{equation}
\J_{A+B}=\ln h_{q}(g_{A}g_{B})=\ln h_{q}\left[H_{q}(e^{\J_{A}})H_{q}(e^{\J_{B}})\right]=\J_{A}+\J_{B}+(1-q)\J_{A}\J_{B} 
\end{equation}

\subsection{Probability}

The probability distributions can be obtained from Eqs. (\ref{pflug}) and (\ref{pbgen})
\begin{eqnarray}
&&\mathcal{P}_{\{X_{s}\}}=\frac{g_{\{X_{j}\} \cup \{X_{s}\}}
\left(1+\frac{y_{s}X_{s}(q-1)}{g_{\{X_{j}\} \cup \{X_{s}\}}^{1-q}}
\right)^{\frac{1}{1-q}}}
{\sum_{\{X_{s}\}}
g_{\{X_{j}\} \cup \{X_{s}\}}
\left(1+\frac{y_{s}X_{s}(q-1)}{g_{\{X_{j}\} \cup \{X_{s}\}}^{1-q}}
\right)^{\frac{1}{1-q}}}=e^{\frac{1}{1-q}\ln \left[\frac{1+\ekx_{\{X_{s}\}}(q-1)}{1+\ekx(q-1)}\right]}
\label{qfluctu} \\
\nonumber \\
\nonumber \\
&&p_{k,\{X_{s}\}}=\frac{\left(1+\frac{y_{s}X_{s}|_{k}(q-1)}{g_{\{X_{j}\} \cup \{X_{s}|_{k}\}}^{1-q}}
\right)^{\frac{1}{1-q}}} {\sum_{k}
\left(1+\frac{y_{s}X_{s}|_{k}(q-1)}{g_{\{X_{j}\} \cup \{X_{s}|_{k}\}}^{1-q}}
\right)^{\frac{1}{1-q}}}=
e^{\frac{1}{1-q}\ln \left[\frac{1+\ekx_{\{X_{s}\}}(q-1)}{\left[1+\ekx(q-1)\right]
\left[1+\J_{\{X_{s}\}}(1-q)\right]}\right]}
\label{probtsal}
\end{eqnarray}
where Eqs. (\ref{genfl}) and (\ref{genfl2} have also been used.
These probability distributions are normalized.

The observed values of extensive non-environment variables can be obtained from Eq.(\ref{genobs})
\begin{equation}
X_{i,obs}=\sum_{X_{i}}\mathcal{P}^{q}_{X_{i}}X_{i}=\sum_{k}\frac{p_{k}^{q}}
{\left[g_{\{X_{j}\} \cup \{X_{i}|_{k}\}}\right]^{1-q}}X_{i}|_{k}
=<X_{i}>_{q} \label{qobs}
\end{equation}
where $<X_{i}>_{q}$ denotes the q-mean value. In general, for a mechanical quantity $A$, we have
\begin{equation}
<A>_{q} = \sum_{k}\frac{p_{k}^{q}}
{\left[g_{\{X_{j}\} \cup \{X_{i}|_{k}\}}\right]^{1-q}}A|_{k} 
= \sum_{k}\frac{ g_{\{X_{j}\} \cup \{y_{i}\} \cup \{X_{i}|_{k}\}}^{q} }
{g_{\{X_{j}\} \cup \{X_{i}|_{k} \}} g_{\{X_{j}\} \cup \{y_{i}\}}^{q}
}A|_{k}
\end{equation}
This q-mean value is different to those suggested by Tsallis et al. \cite{Mendes}. Let us analyze
briefly some of the implications of this result. In particular, we see that in the microcanonical
ensemble the norm is conserved, i.e.
\begin{equation}
<1>_{q} = \sum_{k}\frac{ g_{\{X_{l}\}}^{q} }
{g_{\{X_{l}\}} g_{\{X_{l}\}}^{q}}=1
\end{equation}
The norm is not conserved, in the rest of statistical ensembles. 
Only integrals of motion are properly defined in the microcanonical
ensemble, however, and the lack of normalization in other ensembles is 
not a matter for worrying about here. The average $\sum_{k}p_{k}^{q}A|_{k}/\sum_{k}p_{k}^{q}$
recently considered could be obtained with other definition of probability different to the given above. 
We choose however to maintain the same definition of probability as we given for equilibrium.
Operationally, the mean value defined by us has the same advantages that the so-called
second choice for the mean value proposed by Tsallis: $<A>_{q}=\sum_{k}p_{k}^{q}
A|_{k}$. Furthermore, the conservation of the norm is at least preserved in 
the microcanonical ensemble.

For the entropy, we have also have,
by using Eqs.(\ref{qgenent}), (\ref{qekx}) and (\ref{qobs})
\begin{eqnarray}
&&\J=y_{i}X_{i,obs}-\ekx \nonumber \\
&&=\frac{1}{1-q}\left(\sum_{k}p_{k}^{q}\frac{y_{i}X_{i}|_{k}(1-q)}
{\left[g_{\{X_{j}\} \cup \{X_{i}|_{k}\}}\right]^{1-q}}
+g_{\{X_{j}\} \cup \{y_{i}\}}^{1-q}-1\right) \nonumber \\
&&=\frac{1}{1-q}\left(\sum_{k}\left[p_{k}^{q}
-p_{k}g_{\{X_{j}\} \cup \{y_{i}\}}^{1-q} \right]
+g_{\{X_{j}\} \cup \{y_{i}\}}^{1-q}-1\right) \nonumber \\
&&=\frac{\sum_{k}p_{k}^{q}-1}{1-q}
\end{eqnarray}
which is the Tsallis generalized entropy. This was the starting point for Tsallis
to develop nonextensive statistics.

Einstein's fluctuation formula can be easily deduced for Tsallis statistics by considering
a general vector $\vec{\alpha}$ in Eq.(\ref{qfluctu}) instead of variables $\{X_{s}\}$ and expanding
$\ekx_{\vec{\alpha}}$ to second order in $\vec{\alpha}$
\begin{equation}
\mathcal{P}_{\vec{\alpha}}=\mathcal{P}_{0}e^{\frac{1}{1-q}\ln
\left[\frac{1+(q-1)\ekx_{\vec{\alpha}}}{1+(q-1)\ekx_{0}}\right]}
\end{equation}
We obtain
\begin{equation}
\mathcal{P}_{\vec{\alpha}} \sim \mathcal{P}_{\vec{\alpha}_{0}} e^{-\frac{1}{2}
\frac{\left(\frac{\partial^{2} \ekx_{\vec{\alpha}}}
{\partial \alpha_{k} \partial \alpha_{l}}\right)_{0}}
{1+(q-1)\ekx_{0}}\alpha_{k}\alpha_{l}}
\label{qEinstein}
\end{equation}
If the matrix $\mathbf{G}$ defined above, Eq.(\ref{G}), is used,
the variances can now be calculated from Eq.(\ref{qEinstein}) 
for $\vec{\alpha}$ and $\vec{\lambda}\equiv \mathbf{G}\vec{\alpha}$
following an analogous
procedure to that of Section \ref{fluctu}. These
are given by
\begin{eqnarray}
<\vec{\lambda}\vec{\lambda}>&=&\left[1+(q-1)\ekx_{0} \right]\mathbf{G} \label{qfluctuin} \\
<\vec{\alpha}\vec{\alpha}>&=&\left[1+(q-1)\ekx_{0} \right]\mathbf{G}^{-1} \label{qfluctuex}
\end{eqnarray}
And thus
\begin{equation}
q-1=\frac{\sqrt{<\vec{\lambda}\vec{\lambda}><\vec{\alpha}\vec{\alpha}>}-1}{\ekx_{0}}
\end{equation}
It is noteworthy that in the limit $q \to 1$
all expressions obtained in this Section reduce to those obtained in Sections
\ref{ET}, \ref{prob} and \ref{fluctu}. It must also be noted that the thermodynamic formalism
exposed in Section \ref{charf} does not change. A clear connection between Tsallis statistics
and thermodynamics is thus established and Legendre transform is warranted in the thermodynamic
limit.

The treatment followed embodies Hill thermodynamics and, hence, is stable, allowing for large fluctuations in physical quantities.
The formulation has been based in the concepts of classes of configurations and in the principle of minimum for the dimensionless
(generalized) characteristic thermodynamic potential. Both concepts have a clear meaning in thermostatistics as we have shown.

We conclude remarking that we have formulated a possible generalization of BG thermostatistics in Section 
\ref{generalization}. We call this formalism Squeezed Statistical Mechanics due to the fact that it makes reference
to a squeezed phase space because of the introduction of the squeezing function $h_{q}$. 
In the framework of this alternative formulation we have developed classical equilibrium,
BG statistical mechanics (supposing no squeezing) and a general framework valid for nonequilibrium situations
and complex systems. Among the advantages of the formulation presented here are the compactness and generality of the expressions obtained and
the rationalization of recent statistical mechanics developments in an unified approach.
New insights in these developments have been provided. Furthermore, the general formulation presented
is fully consistent with a generalized kinetic theory previously proposed. Applications to concrete physical systems
of the formalism presented will be given in forthcoming papers. We indicate however that previous applications of Tsallis and Beck-Cohen 
statistics and also Hill's nanothermodynamics can be considered also as applications of the extended theory presented here.
Moreover, the treatment proposed can be extended to other more complex choices for the squeezing function different
to the one-parameter dependent family we have considered in this paper. Linkings to 
the very interesting and important formalisms including SRB distributions and the chaotic hypothesis
are also to be explored.

We are very grateful to Dr. Kaniadakis for bringing our attention to reference \cite{Kan2} 
and suggesting the direct connection of our formalism with his generalized kinetic model.
We want also thank J. A. Manzanares and S. Maf\'{e} for several comments
on the manuscript. One of us (V. Garc\'{\i}a-Morales) wishes also to thank 
support from the Spanish M. E. C. D (grant No. AP2001-3329)
and conversations with J. M. Garc\'{\i}a-Sanchis.

\end{document}